\documentclass[runningheads]{llncs}
\usepackage[T1]{fontenc}
\usepackage{hyperref}
\hypersetup{
  colorlinks=true,
  linkcolor=blue,
  urlcolor=blue,
  citecolor=black,
}
\usepackage{graphicx}
\usepackage{booktabs, multirow} 
\usepackage{amssymb}
\usepackage{soul}
\usepackage{xcolor,colortbl} 
\usepackage{changepage,threeparttable} 
\usepackage{subfigure}
\usepackage[most]{tcolorbox}
\newtcbtheorem{Summary}{\bfseries Takeaways}{enhanced,drop shadow={black!50!white},
  coltitle=black,
  top=0.1in,
  attach boxed title to top left=
  {xshift=1.5em,yshift=-\tcboxedtitleheight/2},
  boxed title style={size=small,colback=gray}
}{summary}
\begin{document}
\title{LUMIA: Linear probing for Unimodal and MultiModal Membership Inference Attacks leveraging internal LLM states}
%
%

\titlerunning{LUMIA: Handling Inference Attacks leveraging internal LLM states}

\author{Luis Ibanez-Lissen\inst{1} \and
Lorena Gonzalez-Manzano\inst{1} \and
Jose Maria de Fuentes\inst{1,2}
\and
Nicolas Anciaux\inst{2} \and
Joaquin Garcia-Alfaro\inst{3}}
\authorrunning{Ibanez-Lissen et al.}
%
\institute{Universidad Carlos III de Madrid, Leganes, Spain  \and
Inria, UVSQ, U. Paris-Saclay, Palaiseau, France \and
SAMOVAR, T\'el\'ecom SudParis, Institut Polytechnique de Paris, Palaiseau, France}
\maketitle              
\begin{abstract}
Large Language Models (LLMs) are increasingly used in a variety of applications, but concerns around  membership inference have grown in parallel. Previous efforts focus on black-to-grey-box models, thus neglecting the potential benefit from internal LLM information. To address this, we propose the use of Linear Probes (LPs) as a method to detect Membership Inference Attacks (MIAs) by examining internal activations of LLMs. Our approach, dubbed LUMIA, applies LPs layer-by-layer to get fine-grained data on the model inner workings. We test this method across several model architectures, sizes and datasets, including unimodal and multimodal tasks. In unimodal MIA, LUMIA achieves an average gain of 15.71\% in Area Under the Curve (AUC)  over previous techniques. Remarkably, LUMIA reaches AUC greater than 60\% in 65.33\% of cases -- an increase of 46.80\% against the state of the art. Furthermore, our approach reveals key insights, such as the model layers where MIAs are most detectable. In multimodal models, LPs indicate that visual inputs can significantly contribute to detect MIAs -- AUC greater than 60\% is reached in 85.90\% of the experiments.

\end{abstract}
\section{Introduction}


Membership Inference Attacks (MIAs) aim to determine whether specific data samples (such as sensitive or copyrighted items) were included in the training set of a Large Language Model (LLM) \cite{yan2024protecting}.

While some researchers argue that it is impossible to prove that MIA are feasible on LLMs~\cite{zhang2024membership}, others try to find methods that maximize the Area Under the Curve (AUC) to get better performance. These efforts tackle the membership inference problem from a black-box perspective (e.g., work in~\cite{kim2024detecting}) or a grey-box one. In the latter, the idea is to set a threshold on the model output that determines whether a sample was part of the training data \cite{sablayrolles2019white}, \cite{carlini2021extracting}, \cite{shi2023detecting}. 


\textbf{Motivation.} The need to create fair and transparent auditing processes for AI systems calls for adopting white-box approaches \cite{yan2024protecting,casper2024black}. With regards to MIA, we hypothesize that the internal model data from member and non-member samples may reveal distinguishing patterns. Specifically, we expect that data corresponding to previously seen (member) texts or images may behave differently from unseen (non-member) data. Only Liu \textit{et al.} \cite{liu2024probing} have approached the membership inference problem in LLMs from this perspective. They apply Linear Probes (LPs) \cite{alain2016understanding} on model activations of a single layer. {\color{black}However, their work is preliminary as there are a number of limitations which are tackled in our work. First, their approach involves fine-tuning the models to ensure that members have been seen. Therefore, results are biased since samples already used in the \textit{pretraining} phase are seen twice. Second, such a fine-tuning is used to create  proxy models for the experimentation, but there is no guarantee on the functional equivalence of the original and the proxy model. Thirdly, they use an input prompt, which simplifies the problem as it narrows the search space. Lastly, they are limited to text-based MIAs, thus excluding multimodal models.} 

\textbf{Contribution.} This paper provides an
insightful analysis of the effectiveness on using internal model data for MIA assessment. The approach, dubbed LUMIA\footnote{Latin term derived from \textit{light}, representing the value of looking inside the model to ascertain how MIAs impact the inner model working. }, uses internal activations of each model layer. LUMIA is directly applied on real-world models and datasets, thus characterizing the ability of LPs to succeed depending on the model and the dataset nature. For completeness, we consider two types of biases, in the same line as  Duan \textit{et al.} \cite{duan2024membership} and Das \textit{et al.} \cite{das2024blind}. As no sample prompts are used and LLMs are requested to perform a variety of tasks, our results are easily generalizable. Interestingly, experiments are not only text-based MIAs, but also multimodal. While the concurrent work by Li \textit{et al.} \cite{li2024membership} has proposed a benchmark for multimodal MIAs, they depend on the model output. Therefore, they limit themselves to tasks that generate long texts. On the contrary, LUMIA is not constrained by the LLM output.    

{\color{black}The research question at stake is --- \textit{To what extent can internal activations of LLMs be used to improve and assess membership inference?}} In this vein, the list of contributions is as follows:
\begin{itemize}
    {\color{black}
    \item We provide a comprehensive study on the suitability of internal activations for assessing MIAs by using linear probes, showing their ability to outperform the state of the art.}
    \item We explore for the first time the impact of the LLM size, the dataset nature and bias and the impact of using deduplicated model versions. 
    \item We analyse the problem of MIAs in multimodal LLMs. {\color{black}We consider a variety of LLM tasks, which has never been tackled to the best of authors knowledge.} 
    \item  Our experimental results are based on 14 textual and seven multimodal datasets and three model families, involving 15 LLM configurations. We release our experimental materials to foster further research in this direction\footnote{\textit{Kindly find at \href{https://github.com/Luisibear98/LUMIA}{https://github.com/Luisibear98/LUMIA} a reduced version of LUMIA, for reviewing purposes.}}. 
\end{itemize}

{
\color{black}
This paper is structured as follows: Section \ref{sec:Preliminaries} provides the necessary background information. Section \ref{sec:foundations} describes the foundations of LUMIA, whereas Section \ref{sec:expe} covers all the experimentation, which is later analyzed in Section \ref{sec:results}. Section \ref{RW} shows the related work. Lastly, Section \ref{sec:conclusion} concludes the paper and points out future work directions.
}

\vspace{-.35cm}

\section{Background}
\label{sec:Preliminaries}
The background on Large Language Models (LLMs) is introduced in Section \ref{sec:llm}. Afterwards, the basics of linear probes are described in Section \ref{sec:lps}.

\vspace{-.35cm}

\subsection{LLMs. Internal data, input data and biases}
\label{sec:llm}

LLMs are transformer-based neural networks \cite{vaswani2017attention}, consisting of tens to hundreds of billions of parameters, and pretrained on vast amounts of  data. Notable examples include models like LLaMA \cite{dubey2024llama} and GPT-4 \cite{achiam2023gpt}. For the interest of this paper, the information processed and stored by the neural network during training and inference is at stake. In the transformer model \cite{vaswani2017attention}, internal model data specifically refers to the activations generated at the output of each transformer block during the feed-forward pass. These activations represent the intermediate state of the model as it processes input data layer by layer. 

\color{black}
A key factor in training these models is ensuring data quality, especially when scraping large corpora. One way to measure data quality is by identifying biases. We hereby describe the two major types of bias \cite{gao2000n}. One way is analyzing N-grams, which are sequences of $n$ consecutive elements (e.g., words in natural language processing). In MIA, N-gram overlap indicates the percentage of N-grams in a non-member sample that also appear in at least one member sample. Thus, higher overlaps imply more similarity across samples  \cite{duan2024llms}. In this proposal, we call this potential source of bias as N-gram bias (NGB).

Another form of bias in MIA arises from dynamic changes in data distribution over time. Thus, members are typically selected before a given date, and non-members are those after that deadline. Das \textit{et al.} \cite{das2024blind} identified this issue, which we refer to as temporal bias (TB).

\vspace{-.35cm}

\subsection{Linear classifier probes}
\label{sec:lps}
Linear Probes (LP) are classifiers (such as Multi-Layer Perceptrons, MLPs) that contribute to deep learning models explainability efforts by providing insights into how the model processes information internally \cite{alain2016understanding}. 

LPs are used to make predictions over the hidden states of the models, trying to predict or identify if some specific information is correctly represented within them.
For LLMs, an LP classifier is typically placed after each layer of the network and takes the hidden states as input $X$ to predict a concept $Y$. 

\section{LUMIA}
\label{sec:foundations}

This section provides the foundations of our proposal. Section~\ref{sec:problemformulation} covers the formulation of the problem and Section~\ref{sec:overview} describes the approach.

\subsection{Problem formulation}
\label{sec:problemformulation}

{\color{black} 

LUMIA (Linear probe-based Utilization of Model Internal Activations), leverages Linear Probes (LPs), lightweight classifiers trained directly on internal activations, i.e. the hidden states generated at each layer during inference. LPs offer an interpretable and efficient means to assess the distribution of membership information across the model's layers. Specifically, we formalize the problem of membership inference using internal activations as follows:

\begin{itemize}
    \item \textbf{Input:} A pre-trained model \( M \), and a set of labeled samples \( S = \{(x_i, y_i) \} \). \( x_i \)=$\lbrace t_1, \cdots, t_k \rbrace$ is the input (text or multimodal text-image pair) formed by minimal data units called tokens $t_i$. \( y_i \in \{0, 1\} \) indicates membership status ($1$ if the sample is a member, $0$ otherwise).
    \item \textbf{Objective:} Train a linear probe \( MLP_l \) for each layer \( l \) of the model \( M \) to classify membership status based on the internal activation \( A_l(x_i) \), where \( A_l(x_i) \) represents the average activation vector at layer \( l \) for all tokens $t_i$ within input \( x_i \).
    \item \textbf{Metric:} Evaluate \( P_l \) using metrics such as Area Under the Curve (AUC) for each layer $l$, and identify the layer \( l^* \) where membership information is most detectable (i.e. where \( P_{l^*} \) achieves the highest AUC).
\end{itemize}

This formulation enables us to explore:
\begin{enumerate}
    \item The distribution and concentration of membership information across different layers of LLMs. 
    \item The comparative effectiveness of LP-based MIAs versus traditional output-based methods.
    \item The influence of factors such as model architecture, size, dataset characteristics, and multimodal inputs on membership inference success.
\end{enumerate}

By rigorously applying this approach, LUMIA aims to advance the understanding of membership inference in LLMs, and establishes internal activations as versatile and powerful tool for MIA assessment.}

\subsection{Description}
\label{sec:overview}

\begin{figure}[!t]
    \centering
\includegraphics[width=\columnwidth]{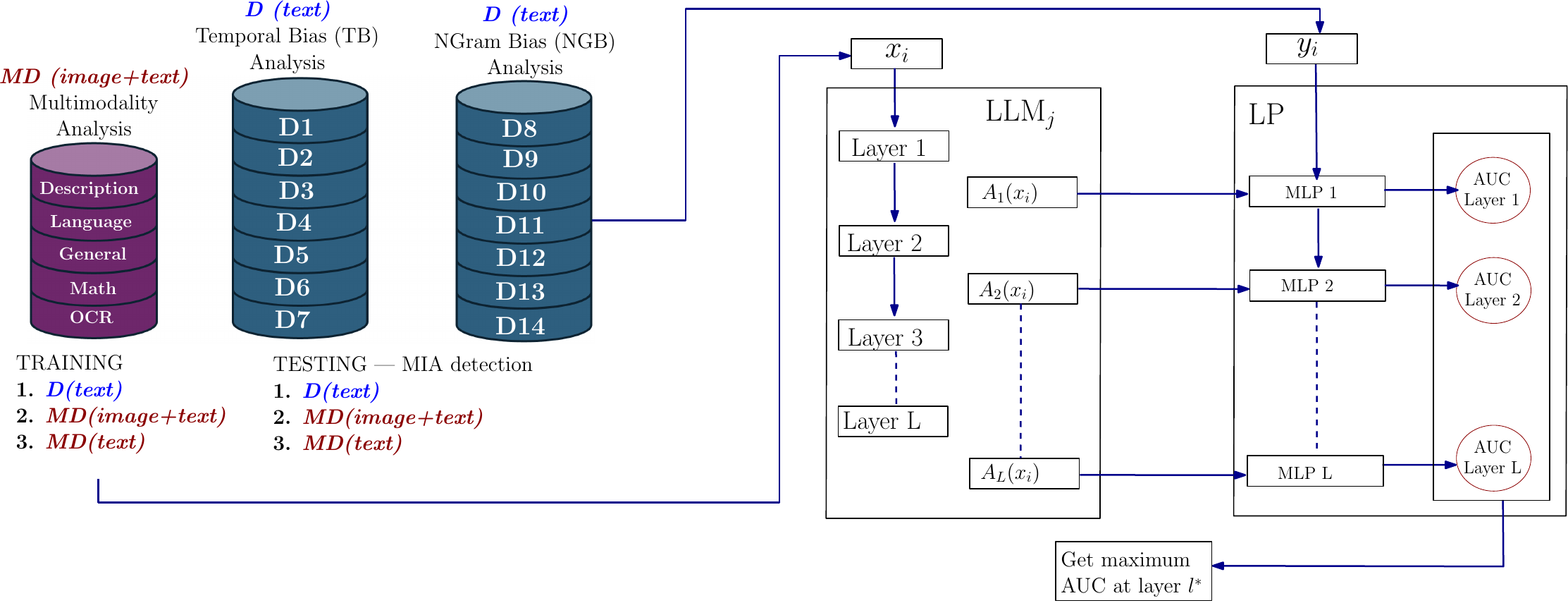}
    \caption{System overview}
    \label{fig:overview}
\end{figure}

Depicted in Figure \ref{fig:overview}, once an LLM is trained with member and non-member samples, internal activations at each layer are input of an LP. LPs are implemented through MLPs (recall Section \ref{sec:lps}), whose output is AUC. LUMIA retrieves the AUC per layer as well as the layer \( l^* \) in which the maximum AUC is achieved.

To ensure the robustness and generalization of this process, unimodal ($D$) and multimodal ($MD$) datasets are applied over multiple LLM (e.g. $LLM_j$), where $D$ provides answers to a general instruction prompt of any type, e.g. make a summary. More specifically, $D$ are used to test the improvement of LUMIA over N-gram bias (NGB), thus studying the benefits of using LPs for MIA attacks with different levels of overlapping among inputs. $D$ are also applied to study the effect of temporal bias (TB).

Given the large variety of LLMs, $MD$ allows analysing MIA attacks once samples composed of image and text are input. A couple of ways to handle multimodality are devised -- training an LLM just with images or with images and text, computing LPs over the resulting activations \( A_l(x_i) \).

\subsubsection{Extracting activation data}
\label{activationprocess}
Activations \( A_l(x_i) \), per layer, capture values for all tokens. First, samples  \( x_i \)  from members and non-members are preprocessed by cropping the text to fit the maximum context length $n$ of the target LLM. Next, a forward pass \cite{rumelhart1986learning} is performed for each sample, during which the hooks capture the activations  $a_i(t_i)$  for each token  \( t_i \) at layer \( l \). Thus, their average is computed as follows:
\begin{equation}
    A_l(x_i) = \frac{1}{n} \sum_{j=1}^n a_l(t_j)
\end{equation}

For unimodal cases, hooks are placed after each transformer layer. In multimodal cases, hooks are positioned after the layers of both the text and visual models.

\section{Experiment design}
\label{sec:expe}

This section describes the design of the experiments to assess LUMIA. Models and datasets are explained on Section \ref{sec:model_data}. Section \ref{sec:eval_metrics} describes the assessment metrics. Finally, Section \ref{sec:expSetting} introduces the experimental settings.

\subsection{Models, datasets and tasks}
\label{sec:model_data}

\noindent \textbf{Unimodal LLMs.} Several models of different sizes are chosen in this study. On the one hand, the Pythia model family \cite{biderman2023pythia}, trained on the Pile dataset \cite{gao2020pile}, with 160M, 1.4B, 2.8B, and 12B of parameters was selected in both their non-deduplicated and deduplicated versions for comparison purposes. Additionally, the GPT-Neo family is also evaluated with 140M, 1.3B, and 2.7B parameters variants. These models are chosen (1) to compare them to other proposals, and (2) because data used for pre-training them is known, being essential to deal with MIA attacks.

\medskip

\noindent \textbf{Unimodal task and datasets.} In line with the state of the art, the LLM processes text to carry out a text-masking causal modeling task. In this vein, datasets used to test the approach are \textit{WikiMIA} \cite{shi2023detecting}, \textit{ArXiv-MIA} \cite{liu2024probing}, \textit{Temporal ArXiv/wiki} \cite{duan2024membership}, \textit{ArXiv-1-month} \cite{meeus2024did}, \textit{Gutenberg} \cite{meeus2024did} and  \textit{Mimir} \cite{duan2024membership}. Note that they have been selected for the sake of comparability with previous works \cite{shi2023detecting,liu2024probing,duan2024membership,meeus2024did}. All datasets, except for Mimir, have already been shown to suffer from TB \cite{das2024blind}. 
Conversely, Mimir suffers from NGB. 

\medskip

\noindent \textbf{Multimodal LLMs.} For the analysis of multimodality, the latest version of the LLava-OneVision model \cite{li2024llava} is applied with 0.5B and 7.6B parameters. These models are chosen since  (1) the data used during its pre-training and fine-tuning is known and, (2) due to available computational resources.

\medskip

\noindent \textbf{Multimodal tasks and datasets.} Linked to this model,  \textit{OneVision-Data}\footnote{\href{https://huggingface.co/datasets/lmms-lab/LLaVA-OneVision-Data}{https://huggingface.co/datasets/lmms-lab/LLaVA-OneVision-Data}, last accessed on January 10, 2025.} dataset is applied. It is composed of a wide range of datasets used to train a multimodal model for multitasking. From this collection, we generate member and non-member samples from datasets that originally provided distinct training, validation, and testing splits. From all datasets, the following are selected -- \textit{Textcaps} \cite{sidorov2020textcaps}, \textit{MathV360k} \cite{shi2024math}, \textit{AOK} \cite{schwenk2022okvqa}, \textit{ChartQA} \cite{masry2022chartqa}, \textit{ScienceQA} \cite{lu2022learn},  \textit{IconQA} \cite{lu2021iconqa} and \textit{Magpie} \cite{xu2024magpie}. They encompass all the modalities and categories of tasks the model can accomplish: General resolution, Doc/Chart/Screen solving, Math/Reasoning, OCR and Language tasks.

\subsection{Metrics. Performance and bias}
\label{sec:eval_metrics}
In line with Duan \textit{et al.} \cite{duan2024membership}, Shi \textit{et al.}\cite{shi2023detecting}, and Carlini \textit{et al.} \cite{carlini2022membership} (and for the sake of comparison), the effectiveness of the detection method is measured with the following metrics:
\begin{itemize}
    \item \textbf{Area Under the ROC Curve (AUC).} It measures the  ability of a classifier to correctly determine a class, 0 or 1, by comparing the true positive rate (power) against
the false positive rate (error) across various thresholds.
A value closer to 1 means better performance. In line with \cite{duan2024llms}, MIA will be considered successful when AUC is higher than 0.6.

AUC is then computed to compare LUMIA against state-of-the-art MIAs, namely \textit{Loss} \cite{yeom2018privacy}, \textit{Reference-based}\cite{sablayrolles2019white}, \textit{Zlib Entropy} \cite{carlini2021extracting} and \textit{Min-k\% Probability} \cite{shi2023detecting}.

\end{itemize}

Concerning text-based bias, only NGB can be measured. In this regard, we use the n-gram length $\mathcal{N}$ and the percentage of overlap $\mathcal{P}$, in line with \cite{duan2024llms}.
 
 In case of multimodality, there are no standard, widely accepted metrics in this regard. Thus, the following image signal processing techniques are computed in members and non members \cite{wang2004image,wang2009mean}:
\begin{itemize}
    \item \textbf{Average Hash variation (HV)}. Images are converted to grey scale and the average value of the pixels is computed. Finally a hash is applied to compare similarity across samples. 
    \item \textbf{Average Structural similarity index measure (SSIM)}. It refers to the perceptual similarity between  images by considering their luminance, contrast, and structural content.

\end{itemize}

\subsection{Experimental settings}
\label{sec:expSetting}

Training was conducted on two NVIDIA consumer GPUs, a RTX 4090 and a RTX 4080, using mix of the Pytorch, Tensorflow frameworks and the Hugging Face library\footnote{\href{https://huggingface.co}{https://huggingface.co}, last accessed on January 10, 2025.}. For both training and validation, all datasets were randomly split in an 80\%-20\% balancing both classes (members and non-members) and repeating three times experiments with different samples. The average of all executions is then computed. 
MLPs models are trained with a learning rate of $1e^{-3}$, using the Adam optimizer \cite{kingma2014adam} over 100 epochs with early stops and dropout regularization. 

For comparison with related work~\cite{shi2023detecting,liu2024probing,duan2024membership,meeus2024did}, we extract 1000 members and 1000 non-members per dataset, except for WikiMIA and ArXiv-MIA. For these cases, we use the provided data: 250 and 400 samples per class, respectively. In the case of multimodality, we also create a joint subset  extracting 100 samples from all datasets, forming a total of 700 members and 700 non-members.

For multimodal configurations, we pick the members and non-members using the IDs provided on the original datasets to avoid contamination between training, validation and testing sets.

Lastly, since the original Magpie setup does not provide image inputs but the model requires both text and image modalities, we pair each text input with a black image to create the necessary input pairs.

\section{Results}
\label{sec:results}

This section presents the results of LUMIA. Firstly, how LP outperforms MIA attacks is analysed (Section~\ref{sec:outperform}), followed by a study  of the impact of target model size (Section~\ref{sec:modelSize}). The influence of potential bias is then explored (Section~\ref{sec:impact_bias}), along with the role of dataset nature (Section~\ref{sec:dataNature}), the effects of data deduplication (Section~\ref{sec:dataDedup}), and the significance of layer depth (Section~\ref{sec:layerDep}).

Results are presented in the form of tables which are used across all sections. For the sake of clarity, Tables~\ref{tab:compProbes} and Table~\ref{tab:general} highlight LUMIA values where AUC is higher than 60\%, while Table~\ref{tab:compmulti} highlights the best values for each setting.

\subsection{Overall effectiveness}
\label{sec:outperform}

\noindent \textbf{Unimodal.} Table \ref{tab:compProbes} and Table \ref{tab:general} summarize the results of our approach versus all the previous proposals (hereinafter \textit{Best SOTA AUC}). LUMIA overtakes previous results on all the cases except in two, which represents an improvement on 174 of the 176 cases (98.86\%). Indeed, our approach provides an average AUC improvement of 15.75\%.  Considering an AUC greater than 0.6 as threshold \cite{duan2024llms}, previous approaches surpass that value on the 44.5\% of the cases while LUMIA reaches that threshold on the 65.33\% of the cases, that is an increment of 46.80\%. 

\begin{table}[!b]
  \centering
  \caption{AUC comparison with State of the art (SOTA) on TB datasets.}
       \scalebox{0.9}{
\begin{tabular}{ccccccc}
\hline
\textbf{Method} & \multicolumn{4}{c}{\textbf{Best SOTA AUC}} & \textbf{Ours} & \textbf{Improvement} \\ \hline
\multicolumn{7}{c}{\textbf{Gutenberg}} \\
 \hline
Document features$^1$ & \multicolumn{4}{c}{0.856} & \multirow{2}{*}{\textbf{0.98}} & 14.49\% \\
Heuristics$^2$  & \multicolumn{4}{c}{0.964} &  & 1.66\% \\ \hline
\multicolumn{7}{c}{\textbf{ArXiv-1 month}} \\ \hline
Document features$^1$  & \multicolumn{4}{c}{0.678} & \multirow{2}{*}{\textbf{0.93}} & 37.17\% \\
Heuristics$^2$  & \multicolumn{4}{c}{0.684} &  & 35.96\% \\ \hline
\multicolumn{7}{c}{\textbf{Temporal wiki}} \\ \hline
Best-Duan$^3$  & \multicolumn{4}{c}{0.796} & \multirow{2}{*}{\textbf{0.93}} & 16.83\% \\
Heuristics$^2$  & \multicolumn{4}{c}{0.799} &  & 16.40\% \\ \hline
\multicolumn{7}{c}{\textbf{Temporal ArXiv 2020-08}} \\ \hline
Best-Duan$^3$  & \multicolumn{4}{c}{0.723} & \multirow{2}{*}{\textbf{0.86}} & 18.32\% \\
Heuristics$^2$  & \multicolumn{4}{c}{0.756} &  & 13.15\% \\ \hline
\multicolumn{7}{c}{\textbf{WikiMIA}} \\ \hline
Min prob$^4$  & \multicolumn{4}{c}{0.839} & \multirow{3}{*}{\textbf{0.99}} & 18.00\% \\
Finetune + probes$^5$ & \multicolumn{4}{c}{0.698} &  & 41.83\% \\
Heuristics$^2$  & \multicolumn{4}{c}{0.987} &  & 0.30\% \\
EM-MIA$^6$ & \multicolumn{4}{c}{0.977} &  & 1.33\% \\ 
ModRényi$^7$ & \multicolumn{4}{c}{0.809} &  & 22.37\% \\ \hline
\multicolumn{7}{c}{\textbf{ArXiv CS}} \\ \hline
Finetune + probes$^5$  & \multicolumn{4}{c}{0.673} &\textbf{ 0.842 }& 25.11\% \\ \hline
\multicolumn{7}{c}{\textbf{ArXiv Math}} \\ \hline
Finetune + probes$^5$  & \multicolumn{4}{c}{0.574} & \textbf{0.646} & 12.54\% \\ \hline
\hline
\textbf{Average improvement}  & \multicolumn{4}{c}{} &  & 18.00\% \\ \hline

\end{tabular}}
  \label{tab:compProbes}%
  \begin{minipage}{8cm}%
 \small{ Meeus \textit{et al. } \cite{meeus2024did}$^1$ Das \textit{et al. } \cite{das2024blind}$^2$  Duan \textit{et al. } \cite{meeus2024did}$^3$ Shi \textit{et al. } \cite{shi2023detecting}$^4$ Liu \textit{et al. } \cite{liu2024probing}$^5$   Kim \textit{et al. }\cite{kim2024detecting}$^6$    Li \textit{et al. }\cite{li2024membership}$^7$   
    }
  
  \end{minipage}%

\end{table}

\begin{table*}[!hbt]
  \caption{AUC comparison against Duan \textit{et al.} \cite{duan2024membership} for NGB datasets }
     \scalebox{0.56}{
\begin{tabular}{ccccc|ccc|ccc}
\hline
\multicolumn{1}{l}{} & \multicolumn{9}{c}{\textbf{Pythia Dedup 12B}} & \multicolumn{1}{l}{} \\ \hline
\textbf{Dataset} & \textbf{MIA} & \textbf{ \textbf{$\mathcal{N}=13$  $\mathcal{P}=0.8$} \cite{duan2024membership}} & \textbf{Ours} & \textbf{Improvement} & \textbf{ \textbf{$\mathcal{N}=13$  $\mathcal{P}=0.2$} \cite{duan2024membership}} & \textbf{Ours} & \textbf{Improvement} & \textbf{ \textbf{$\mathcal{N}=7$  $\mathcal{P}=0.2$} \cite{duan2024membership}} & \textbf{Ours} & \textbf{Improvement} \\ \hline
\multirow{4}{*}{Wikipedia} & LOSS & 0.516 & \multirow{4}{*}{0.570} & 10.47\% & 0.545 & \multirow{4}{*}{0.590} & 8.26\% & 0.666 & \multirow{4}{*}{\textbf{0.690}} & 3.60\% \\
 & Ref & 0.578 &  & -1.38\% & 0.590 &  & 0.00\% & 0.677 &  & 1.92\% \\
 & min-k & 0.517 &  & 10.25\% & 0.562 &  & 4.98\% & 0.644 &  & 7.14\% \\
 & zlib & 0.524 &  & 8.78\% & 0.543 &  & 8.66\% & 0.631 &  & 9.35\% \\ \hline
\multirow{4}{*}{Github} & LOSS & 0.678 & \multirow{4}{*}{\textbf{0.770}} & 13.57\% & 0.802 & \multirow{4}{*}{\textbf{0.910}} & 13.47\% & 0.878 & \multirow{4}{*}{\textbf{0.930}} & 5.92\% \\
 & Ref & 0.559 &  & 37.75\% & 0.615 &  & 47.97\% & 0.615 &  & 51.22\% \\
 & min-k & 0.683 &  & 12.74\% & 0.830 &  & 9.64\% & 0.890 &  & 4.49\% \\
 & zlib & 0.690 &  & 11.59\% & 0.829 &  & 9.77\% & 0.908 &  & 2.42\% \\ \hline
\multirow{4}{*}{Pubmed} & LOSS & 0.506 & \multirow{4}{*}{0.580} & 14.62\% & 0.534 & \multirow{4}{*}{0.570} & 6.74\% & 0.780 & \multirow{4}{*}{\textbf{0.980}} & 25.64\% \\
 & Ref & 0.559 &  & 3.76\% & 0.573 &  & -0.52\% & 0.595 &  & 64.71\% \\
 & min-k & 0.512 &  & 13.28\% & 0.542 &  & 5.17\% & 0.792 &  & 23.74\% \\
 & zlib & 0.506 &  & 14.62\% & 0.537 &  & 6.15\% & 0.772 &  & 26.94\% \\ \hline
\multirow{4}{*}{Pile CC} & LOSS & 0.516 & \multirow{4}{*}{\textbf{0.600}} & 16.28\% & 0.534 & \multirow{4}{*}{\textbf{0.601}} & 12.55\% & 0.574 & \multirow{4}{*}{\textbf{0.660}} & 14.98\% \\
 & Ref & 0.582 &  & 3.09\% & 0.593 &  & 1.35\% & 0.644 &  & 2.48\% \\
 & min-k & 0.521 &  & 15.16\% & 0.539 &  & 11.50\% & 0.578 &  & 14.19\% \\
 & zlib & 0.517 &  & 16.05\% & 0.542 &  & 10.89\% & 0.560 &  & 17.86\% \\ \hline
\multirow{4}{*}{ArXiv} & LOSS & 0.527 & \multirow{4}{*}{0.577} & 9.46\% & 0.573 & \multirow{4}{*}{\textbf{0.606}} & 5.76\% & 0.787 & \multirow{4}{*}{\textbf{0.800}} & 1.65\% \\
 & Ref & 0.555 &  & 3.94\% & 0.584 &  & 3.77\% & 0.715 &  & 11.89\% \\
 & min-k & 0.530 &  & 8.84\% & 0.566 &  & 7.07\% & 0.734 &  & 8.99\% \\
 & zlib & 0.521 &  & 10.72\% & 0.565 &  & 7.26\% & 0.780 &  & 2.56\% \\ \hline \hline 
\multirow{4}{*}{DM\_math} & LOSS & 0.485 & \multirow{4}{*}{\textbf{0.600}} & 23.71\% & 0.673 & \multirow{4}{*}{\textbf{0.746}} & 10.79\% & 0.921 & \multirow{4}{*}{\textbf{0.950}} & 3.15\% \\
 & Ref & 0.514 &  & 16.73\% & 0.443 &  & 68.31\% & 0.414 &  & 129.47\% \\
 & min-k & 0.493 &  & 21.70\% & 0.650 &  & 14.71\% & 0.927 &  & 2.48\% \\
 & zlib & 0.481 &  & 24.74\% & 0.643 &  & 15.96\% & 0.805 &  & 18.01\% \\ \hline
\multirow{4}{*}{Hackernews} & LOSS & 0.512 & \multirow{4}{*}{0.584} & 14.01\% & 0.526 & \multirow{4}{*}{0.594} & 12.91\% & 0.604 & \multirow{4}{*}{\textbf{0.690}} & 14.24\% \\
 & Ref & 0.549 &  & 6.33\% & 0.553 &  & 7.40\% & 0.570 &  & 21.05\% \\
 & min-k & 0.526 &  & 10.97\% & 0.533 &  & 11.43\% & 0.585 &  & 17.95\% \\
 & zlib & 0.507 &  & 15.13\% & 0.524 &  & 13.34\% & 0.592 &  & 16.55\% \\ \hline
\hline
\multicolumn{11}{c}{\textbf{GPT-Neo 2.7B}} \\ \hline
\textbf{Dataset} & \textbf{MIA} & \textbf{ \textbf{$\mathcal{N}=13$  $\mathcal{P}=0.8$}} & \textbf{Ours } & \textbf{Improvement} & \textbf{ \textbf{$\mathcal{N}=13$  $\mathcal{P}=0.2$}} & \textbf{Ours} & \textbf{Improvement} & \textbf{ \textbf{$\mathcal{N}=7$  $\mathcal{P}=0.2$}} & \textbf{Ours } & \textbf{Improvement} \\ \hline
\multirow{4}{*}{Wikipedia} & LOSS & 0.513 & \multirow{4}{*}{0.584} & 13.99\% & 0.537 & \multirow{4}{*}{0.58} & 8.01\% & 0.650 & \multirow{4}{*}{\textbf{0.650}} & 0.00\% \\
 & Ref & 0.545 &  & 7.29\% & 0.572 &  & 1.40\% & 0.650 &  & 0.00\% \\
 & min-k & 0.513 &  & 13.99\% & 0.543 &  & 6.81\% & 0.644 &  & 0.93\% \\
 & zlib & 0.519 &  & 12.67\% & 0.535 &  & 8.41\% & 0.623 &  & 4.33\% \\ \hline
\multirow{4}{*}{Github} & LOSS & 0.699 & \multirow{4}{*}{\textbf{0.772}} & 10.53\% & 0.770 & \multirow{4}{*}{\textbf{0.85}} & 10.39\% & 0.878 & \multirow{4}{*}{\textbf{0.940}} & 7.06\% \\
 & Ref & 0.570 &  & 35.55\% & 0.549 &  & 54.83\% & 0.615 &  & 52.85\% \\
 & min-k & 0.700 &  & 10.38\% & 0.802 &  & 5.99\% & 0.890 &  & 5.62\% \\
 & zlib & 0.710 &  & 8.82\% & 0.771 &  & 10.25\% & 0.908 &  & 3.52\% \\ \hline
\multirow{4}{*}{Pubmed} & LOSS & 0.490 & \multirow{4}{*}{0.566} & 15.55\% & 0.498 & \multirow{4}{*}{0.55} & 10.44\% & 0.799 & \multirow{4}{*}{\textbf{0.910}} & 13.89\% \\
 & Ref & 0.507 &  & 11.68\% & 0.507 &  & 8.48\%& 0.786 &  & 15.78\% \\
 & min-k & 0.500 &  & 13.24\% & 0.501 &  & 9.78\% & 0.792 &  & 14.90\% \\
 & zlib & 0.499 &  & 13.47\% & 0.499 &  & 10.22\% & 0.786 &  & 15.78\% \\ \hline
\multirow{4}{*}{Pile CC} & LOSS & 0.500 & \multirow{4}{*}{0.587} & 17.48\% & 0.500 & \multirow{4}{*}{0.59} & 17.91\% & 0.553 & \multirow{4}{*}{\textbf{0.640}} & 15.73\% \\
 & Ref & 0.530 &  & 10.83\% & 0.530 &  & 11.32\% & 0.575 &  & 11.30\% \\
 & min-k & 0.500 &  & 17.48\% & 0.507 &  & 16.37\% & 0.549 &  & 16.58\% \\
 & zlib & 0.500 &  & 17.48\% & 0.505 &  & 16.83\% & 0.540 &  & 18.52\% \\ \hline
\multirow{4}{*}{ArXiv} & LOSS & 0.510 & \multirow{4}{*}{0.586} & 14.92\% & 0.515 & \multirow{4}{*}{0.59} & 14.56\% & 0.790 & \multirow{4}{*}{\textbf{0.860}} & 8.86\% \\
 & Ref & 0.520 &  & 12.71\% & 0.517 &  & 14.12\% & 0.718 &  & 19.78\% \\
 & min-k & 0.517 &  & 13.36\% & 0.519 &  & 13.68\% & 0.760 &  & 13.16\% \\
 & zlib & 0.510 &  & 14.92\% & 0.510 &  & 15.69\% & 0.784 &  & 9.69\% \\ \hline
\multirow{4}{*}{DM\_math} & LOSS & 0.485 & \multirow{4}{*}{0.560} & 15.46\% & 0.676 & \multirow{4}{*}{\textbf{0.75}} & 10.95\% & 0.930 & \multirow{4}{*}{\textbf{1.00}} & 7.53\% \\
 & Ref & 0.509 &  & 10.02\% & 0.435 &  & 72.41\% & 0.502 &  & 99.20\% \\
 & min-k & 0.492 &  & 13.82\% & 0.655 &  & 14.50\% & 0.933 &  & 7.18\% \\
 & zlib & 0.481 &  & 16.42\% & 0.647 &  & 15.92\% & 0.812 &  & 23.15\% \\ \hline
\multirow{4}{*}{Hackernews} & LOSS & 0.502 & \multirow{4}{*}{0.590} & 17.53\% & 0.516 & \multirow{4}{*}{\textbf{0.60}} & 16.28\% & 0.592 & \multirow{4}{*}{\textbf{0.630}} & 6.42\% \\
 & Ref & 0.512 &  & 15.23\% & 0.515 &  & 16.50\% & 0.525 &  & 20.00\% \\
 & min-k & 0.517 &  & 14.12\% & 0.525 &  & 14.29\% & 0.572 &  & 10.14\% \\
 & zlib & 0.502 &  & 17.53\% & 0.519 &  & 15.61\% & 0.587 &  & 7.33\% \\ \hline \hline
 \multicolumn{2}{c}{\begin{tabular}[c]{@{}c@{}}\textbf{Average Improvement}\end{tabular}} & \multicolumn{2}{c}{} & \multicolumn{1}{c|}{11.96\%}  & \multicolumn{2}{c}{} & \multicolumn{1}{c|}{13.93\%} & \multicolumn{2}{c}{} &\multicolumn{1}{c}{17.03\%} \\ \hline
\end{tabular}
}
\label{tab:general}
\end{table*}

\begin{table}[!t]
\centering
\caption{LUMIA AUC results in multimodal models. Notice highlighted best values between models across modalities }
\label{tab:compmulti}
\resizebox{10cm}{!}{%
\begin{tabular}{ccc|cc}
\hline
\multicolumn{1}{l}{\textbf{}} & \multicolumn{2}{c|}{\textbf{Textcaps (General OCR)}} & \multicolumn{2}{c}{\textbf{AOK (General)}} \\ \hline
\textbf{Params} & \textbf{Modality} & \textbf{Best AUC} & \textbf{Modality} & \textbf{Best AUC} \\ \hline
\multirow{2}{*}{0.5B} & Textual + visual & 0.540 & Textual + visual & 0.697 \\
 & Visual & 0.604 & Visual & \textbf{0.735} \\ \hline
\multirow{2}{*}{7B} & Textual + visual & \textbf{0.601} & Textual + visual & \textbf{0.697} \\
 & Visual & \textbf{0.618} & Visual & 0.707 \\ \hline
\multicolumn{1}{l}{} & \multicolumn{2}{c|}{\textbf{ScienceQA (General)}} & \multicolumn{2}{c}{\textbf{ChartQA (Doc/chart/screen)}} \\ \hline
\multirow{2}{*}{0.5B} & Textual + visual & 0.970 & Textual + visual & \textbf{0.694} \\
 & Visual & \textbf{0.806} & Visual & 0.638 \\ \hline
\multirow{2}{*}{7B} & Textual + visual & \textbf{0.990} & Textual + visual & 0.682 \\
 & Visual & 0.802 & Visual & \textbf{0.691} \\ \hline
\multicolumn{1}{l}{} & \multicolumn{2}{c|}{\textbf{magpie (Language)}} & \multicolumn{2}{c}{\textbf{iconqa (General)}} \\ \hline
\multirow{2}{*}{0.5B} & Textual + visual &\textbf{ 0.572} & Textual + visual & 0.869 \\
 & Visual & 0.510 & Visual & 0.809 \\ \hline
\multirow{2}{*}{7B} & Textual + visual & 0.552 & Textual + visual & \textbf{0.903} \\
 & Visual & \textbf{0.520} & Visual & \textbf{0.828} \\ \hline
\multicolumn{1}{l}{} & \multicolumn{2}{c|}{\textbf{Join}} & \multicolumn{2}{c}{\textbf{MathV360k (Math)}} \\ \hline
\multirow{2}{*}{0.5B} & Textual + visual & 0.624 & Textual + visual & 0.599 \\
 & Visual & 0.670 & Visual & 0.584 \\ \hline
\multirow{2}{*}{7B} & Textual + visual & \textbf{0.634} & Textual + visual & \textbf{0.660} \\
 & Visual & \textbf{0.673} & Visual & \textbf{0.629} \\ \hline
\end{tabular}
}
    \begin{minipage}{8cm}%
 \small{\textit{Textual + Visual}: Activations extracted from LLM part.
 
 \textit{Visual}: Activations extracted from Visual encoder part.}

  \end{minipage}%
\end{table}

\medskip

\noindent \textbf{Multimodal.} Table \ref{tab:compmulti} shows the results for the multimodal configurations. All of them, except Magpie, achieve AUC greater than 0.6, suggesting that multimodality may be adding additional information useful for detecting MIA. Magpie reaches an AUC of 0.57, probably because it is the only text-only dataset. When making predictions over a joint dataset, the AUC remains above 0.60 which points out that even when mixing information and modalities, LPs find patterns across activations to define membership. Globally speaking, 85.9\% of cases achieve AUC greater than 0.6, demonstrating better performance as compared to unimodal setups, which meet this threshold in 65.33\% of configurations.

\begin{Summary}{}{firstsummary}
\textit{Unimodal:} We improve 15.71\% AUC on average.  AUC>0.6 is reached by LUMIA 46.8\% more often as compared to the state-of-the-art.

\textit{Multimodal:} No previous work to compare. Among the 7 datasets considered, LUMIA achieves AUC>0.6 in all but one dataset, resulting in 85.9\% of the results exceeding this threshold.

\end{Summary}

\subsection{Impact of model size}
\label{sec:modelSize}

\begin{table}
\centering
\caption{Pythia family. AUC per model size with/without deduplication}
\resizebox{.86\columnwidth}{!}{%
\begin{tabular}{ccccccc|cc}
\hline
\multicolumn{1}{l}{\textbf{}} & \multicolumn{6}{c|}{\textbf{NGB}} & \multicolumn{2}{c}{\textbf{TB}}\\ \hline
\multicolumn{1}{l}{\textbf{}} & \multicolumn{3}{c|}{\textbf{Pythia dedup}} & \multicolumn{3}{c|}{\textbf{Pythia non-dedup}} & \multicolumn{1}{c|}{\textbf{Pythia dedup}} & \textbf{Pythia non-dedup} \\ \hline

\multicolumn{1}{l}{} & \multicolumn{6}{c|}{\textbf{Wikipedia}} & \multicolumn{2}{c}{} \\ \cline{1-7}
\textbf{Params} & \textbf{$\mathcal{N}=13$  $\mathcal{P}=0.8$} & \textbf{$\mathcal{N}=13$  $\mathcal{P}=0.2$} & \multicolumn{1}{c|}{\textbf{$\mathcal{N}=7$  $\mathcal{P}=0.2$}} & \textbf{$\mathcal{N}=13$  $\mathcal{P}=0.8$} & \textbf{$\mathcal{N}=13$  $\mathcal{P}=0.2$} & \textbf{$\mathcal{N}=7$  $\mathcal{P}=0.2$} \ & \multicolumn{2}{c}{\multirow{-2}{*}{\textbf{Gutenberg}}}  \\ \hline 
70m & \cellcolor[HTML]{FFFFFF}0.520 & \cellcolor[HTML]{FEFEFE}0.551 & \multicolumn{1}{c|}{\cellcolor[HTML]{DADADA}0.653} & \cellcolor[HTML]{F7F7F7}0.570 & \cellcolor[HTML]{FFFFFF}0.538 & \cellcolor[HTML]{DADADA}0.651 & \multicolumn{1}{c|}{\cellcolor[HTML]{FFFFFF}0.949} & \cellcolor[HTML]{F1F1F1}0.960 \\
160M & \cellcolor[HTML]{F8F8F8}0.568 & \cellcolor[HTML]{FFFFFF}0.546 & \multicolumn{1}{c|}{\cellcolor[HTML]{D7D7D7}0.660} & \cellcolor[HTML]{FFFFFF}0.546 & \cellcolor[HTML]{F0F0F0}0.589 & \cellcolor[HTML]{D1D1D1}0.676 & \multicolumn{1}{c|}{\cellcolor[HTML]{F1F1F1}0.960} & \cellcolor[HTML]{F1F1F1}0.960 \\
1.4B & \cellcolor[HTML]{FCFCFC}0.557 & \cellcolor[HTML]{F1F1F1}0.586 & \multicolumn{1}{c|}{\cellcolor[HTML]{D1D1D1}0.676} & \cellcolor[HTML]{F9F9F9}0.564 & \cellcolor[HTML]{FBFBFB}0.558 & \cellcolor[HTML]{D6D6D6}0.663 & \multicolumn{1}{c|}{\cellcolor[HTML]{E3E3E3}0.970} & \cellcolor[HTML]{D5D5D5}0.980 \\
2.8B & \cellcolor[HTML]{F4F4F4}0.580 & \cellcolor[HTML]{F6F6F6}0.572 & \multicolumn{1}{c|}{\cellcolor[HTML]{CFCFCF}0.682} & \cellcolor[HTML]{FAFAFA}0.562 & \cellcolor[HTML]{FCFCFC}0.557 & \cellcolor[HTML]{CECECE}0.685 & \multicolumn{1}{c|}{\cellcolor[HTML]{E3E3E3}0.970} & \cellcolor[HTML]{D5D5D5}0.980 \\
12B & \cellcolor[HTML]{F7F7F7}0.570 & \cellcolor[HTML]{F0F0F0}0.590 & \multicolumn{1}{c|}{\cellcolor[HTML]{CCCCCC}0.690} & \cellcolor[HTML]{F7F7F7}0.570 & \cellcolor[HTML]{F4F4F4}0.580 & \cellcolor[HTML]{F0F0F0}0.590 & \multicolumn{1}{c|}{\cellcolor[HTML]{CCCCCC}0.987} & \cellcolor[HTML]{CFCFCF}0.985 \\ \hline
\multicolumn{1}{l}{} & \multicolumn{6}{c|}{\textbf{Github}} & \multicolumn{2}{c}{\textbf{ArXiv-1 month}} \\ \hline
70m & \cellcolor[HTML]{FCFCFC}0.743 & \cellcolor[HTML]{EAEAEA}0.823 & \multicolumn{1}{c|}{\cellcolor[HTML]{D3D3D3}0.922} & \cellcolor[HTML]{FFFFFF}0.732 & \cellcolor[HTML]{E8E8E8}0.832 & \cellcolor[HTML]{DDDDDD}0.881 & \multicolumn{1}{c|}{\cellcolor[HTML]{FFFFFF}0.775} & \cellcolor[HTML]{F7F7F7}0.806 \\
160M & \cellcolor[HTML]{FFFFFF}0.729 & \cellcolor[HTML]{E3E3E3}0.853 & \multicolumn{1}{c|}{\cellcolor[HTML]{E1E1E1}0.863} & \cellcolor[HTML]{FDFDFD}0.741 & \cellcolor[HTML]{DEDEDE}0.874 & \cellcolor[HTML]{D1D1D1}0.931 & \multicolumn{1}{c|}{\cellcolor[HTML]{E9E9E9}0.860} & \cellcolor[HTML]{E7E7E7}0.870 \\
1.4B & \cellcolor[HTML]{F7F7F7}0.767 & \cellcolor[HTML]{E0E0E0}0.868 & \multicolumn{1}{c|}{\cellcolor[HTML]{D0D0D0}0.935} & \cellcolor[HTML]{F2F2F2}0.786 & \cellcolor[HTML]{DEDEDE}0.875 & \cellcolor[HTML]{D3D3D3}0.924 & \multicolumn{1}{c|}{\cellcolor[HTML]{DFDFDF}0.900} & \cellcolor[HTML]{D9D9D9}0.920 \\
2.8B & \cellcolor[HTML]{FAFAFA}0.754 & \cellcolor[HTML]{DFDFDF}0.868 & \multicolumn{1}{c|}{\cellcolor[HTML]{CCCCCC}0.951} & \cellcolor[HTML]{F7F7F7}0.767 & \cellcolor[HTML]{D6D6D6}0.911 & \cellcolor[HTML]{D0D0D0}0.933 & \multicolumn{1}{c|}{\cellcolor[HTML]{D9D9D9}0.920} & \cellcolor[HTML]{D9D9D9}0.920 \\
12B & \cellcolor[HTML]{F6F6F6}0.770 & \cellcolor[HTML]{D6D6D6}0.910 & \multicolumn{1}{c|}{\cellcolor[HTML]{D1D1D1}0.930} & \cellcolor[HTML]{F8F8F8}0.760 & \cellcolor[HTML]{E8E8E8}0.830 & \cellcolor[HTML]{DFDFDF}0.870 & \multicolumn{1}{c|}{\cellcolor[HTML]{EEEEEE}0.843} & \cellcolor[HTML]{EAEAEA}0.856 \\ \hline
\multicolumn{1}{l}{} & \multicolumn{6}{c|}{\textbf{Pile CC}} & \multicolumn{2}{c}{\textbf{Temporal wiki}} \\ \hline
70m & \cellcolor[HTML]{FFFFFF}0.521 & \cellcolor[HTML]{F7F7F7}0.544 & \multicolumn{1}{c|}{\cellcolor[HTML]{E7E7E7}0.587} & \cellcolor[HTML]{F1F1F1}0.561 & \cellcolor[HTML]{F4F4F4}0.552 & \cellcolor[HTML]{E3E3E3}0.599 & \multicolumn{1}{c|}{\cellcolor[HTML]{FDFDFD}0.865} & \cellcolor[HTML]{FFFFFF}0.860 \\
160M & \cellcolor[HTML]{FCFCFC}0.531 & \cellcolor[HTML]{EFEFEF}0.566 & \multicolumn{1}{c|}{\cellcolor[HTML]{E1E1E1}0.603} & \cellcolor[HTML]{F1F1F1}0.562 & \cellcolor[HTML]{F3F3F3}0.554 & \cellcolor[HTML]{DFDFDF}0.611 & \multicolumn{1}{c|}{\cellcolor[HTML]{F5F5F5}0.880} & \cellcolor[HTML]{E4E4E4}0.910 \\
1.4B & \cellcolor[HTML]{EEEEEE}0.569 & \cellcolor[HTML]{ECECEC}0.573 & \multicolumn{1}{c|}{\cellcolor[HTML]{D3D3D3}0.642} & \cellcolor[HTML]{F3F3F3}0.554 & \cellcolor[HTML]{E7E7E7}0.588 & \cellcolor[HTML]{D6D6D6}0.635 & \multicolumn{1}{c|}{\cellcolor[HTML]{E4E4E4}0.910} & \cellcolor[HTML]{DADADA}0.930 \\
2.8B & \cellcolor[HTML]{E6E6E6}0.590 & \cellcolor[HTML]{EDEDED}0.571 & \multicolumn{1}{c|}{\cellcolor[HTML]{DCDCDC}0.618} & \cellcolor[HTML]{EFEFEF}0.566 & \cellcolor[HTML]{E7E7E7}0.587 & \cellcolor[HTML]{D6D6D6}0.634 & \multicolumn{1}{c|}{\cellcolor[HTML]{DADADA}0.929} & \cellcolor[HTML]{DADADA}0.930 \\
12B & \cellcolor[HTML]{E3E3E3}0.600 & \cellcolor[HTML]{EEEEEE}0.570 & \multicolumn{1}{c|}{\cellcolor[HTML]{CCCCCC}0.660} & \cellcolor[HTML]{DFDFDF}0.610 & \cellcolor[HTML]{EEEEEE}0.570 & \cellcolor[HTML]{CCCCCC}0.660 & \multicolumn{1}{c|}{\cellcolor[HTML]{D1D1D1}0.945} & \cellcolor[HTML]{CCCCCC}0.954 \\ \hline
\multicolumn{1}{l}{} & \multicolumn{6}{c|}{\textbf{DM\_math}} & \multicolumn{2}{c}{\textbf{Temporal ArXiv}} \\ \hline
70m & \cellcolor[HTML]{FFFFFF}0.502 & \cellcolor[HTML]{E7E7E7}0.743 & \multicolumn{1}{c|}{\cellcolor[HTML]{D1D1D1}0.959} & \cellcolor[HTML]{FDFDFD}0.525 & \cellcolor[HTML]{E9E9E9}0.721 & \cellcolor[HTML]{D0D0D0}0.963 & \multicolumn{1}{c|}{\cellcolor[HTML]{FFFFFF}0.710} & \cellcolor[HTML]{F4F4F4}0.733 \\
160M & \cellcolor[HTML]{FAFAFA}0.557 & \cellcolor[HTML]{E9E9E9}0.718 & \multicolumn{1}{c|}{\cellcolor[HTML]{D4D4D4}0.927} & \cellcolor[HTML]{FDFDFD}0.526 & \cellcolor[HTML]{EBEBEB}0.706 & \cellcolor[HTML]{CECECE}0.981 & \multicolumn{1}{c|}{\cellcolor[HTML]{FAFAFA}0.720} & \cellcolor[HTML]{FAFAFA}0.720 \\
1.4B & \cellcolor[HTML]{FAFAFA}0.558 & \cellcolor[HTML]{E5E5E5}0.758 & \multicolumn{1}{c|}{\cellcolor[HTML]{D2D2D2}0.943} & \cellcolor[HTML]{FAFAFA}0.559 & \cellcolor[HTML]{E7E7E7}0.745 & \cellcolor[HTML]{CECECE}0.989 & \multicolumn{1}{c|}{\cellcolor[HTML]{E6E6E6}0.760} & \cellcolor[HTML]{EBEBEB}0.750 \\
2.8B & \cellcolor[HTML]{FAFAFA}0.554 & \cellcolor[HTML]{E8E8E8}0.730 & \multicolumn{1}{c|}{\cellcolor[HTML]{CDCDCD}0.995} & \cellcolor[HTML]{F9F9F9}0.570 & \cellcolor[HTML]{E7E7E7}0.741 & \cellcolor[HTML]{D0D0D0}0.961 & \multicolumn{1}{c|}{\cellcolor[HTML]{EBEBEB}0.750} & \cellcolor[HTML]{E6E6E6}0.760 \\
12B & \cellcolor[HTML]{F5F5F5}0.600 & \cellcolor[HTML]{E7E7E7}0.746 & \multicolumn{1}{c|}{\cellcolor[HTML]{CCCCCC}1.000} & \cellcolor[HTML]{F4F4F4}0.610 & \cellcolor[HTML]{F0F0F0}0.650 & \cellcolor[HTML]{CFCFCF}0.980 & \multicolumn{1}{c|}{\cellcolor[HTML]{CCCCCC}0.810} & \cellcolor[HTML]{D3D3D3}0.797 \\ \hline
\multicolumn{1}{l}{} & \multicolumn{6}{c|}{\textbf{Hackernews}} & \multicolumn{2}{c}{\textbf{WikiMIA}} \\ \hline
70m & \cellcolor[HTML]{F7F7F7}0.580 & \cellcolor[HTML]{F2F2F2}0.595 & \multicolumn{1}{c|}{\cellcolor[HTML]{F6F6F6}0.582} & \cellcolor[HTML]{FCFCFC}0.566 & \cellcolor[HTML]{FDFDFD}0.563 & \cellcolor[HTML]{E4E4E4}0.633 & \multicolumn{1}{c|}{\cellcolor[HTML]{FFFFFF}0.970} & \cellcolor[HTML]{E6E6E6}0.980 \\
160M & \cellcolor[HTML]{F2F2F2}0.593 & \cellcolor[HTML]{F8F8F8}0.577 & \multicolumn{1}{c|}{\cellcolor[HTML]{EBEBEB}0.613} & \cellcolor[HTML]{FCFCFC}0.564 & \cellcolor[HTML]{FCFCFC}0.565 & \cellcolor[HTML]{E1E1E1}0.642 & \multicolumn{1}{c|}{\cellcolor[HTML]{FFFFFF}0.970} & \cellcolor[HTML]{E6E6E6}0.980 \\
1.4B & \cellcolor[HTML]{FFFFFF}0.556 & \cellcolor[HTML]{F8F8F8}0.576 & \multicolumn{1}{c|}{\cellcolor[HTML]{E5E5E5}0.630} & \cellcolor[HTML]{F2F2F2}0.594 & \cellcolor[HTML]{FBFBFB}0.570 & \cellcolor[HTML]{DFDFDF}0.647 & \multicolumn{1}{c|}{\cellcolor[HTML]{FFFFFF}0.970} & \cellcolor[HTML]{CCCCCC}0.990 \\
2.8B & \cellcolor[HTML]{F7F7F7}0.579 & \cellcolor[HTML]{F9F9F9}0.576 & \multicolumn{1}{c|}{\cellcolor[HTML]{EBEBEB}0.614} & \cellcolor[HTML]{F3F3F3}0.591 & \cellcolor[HTML]{F7F7F7}0.580 & \cellcolor[HTML]{E1E1E1}0.643 & \multicolumn{1}{c|}{\cellcolor[HTML]{E6E6E6}0.980} & \cellcolor[HTML]{CCCCCC}0.990 \\
12B & \cellcolor[HTML]{F6F6F6}0.584 & \cellcolor[HTML]{F2F2F2}0.594 & \multicolumn{1}{c|}{\cellcolor[HTML]{CCCCCC}0.701} & \cellcolor[HTML]{F0F0F0}0.600 & \cellcolor[HTML]{FEFEFE}0.560 & \cellcolor[HTML]{F7F7F7}0.580 & \multicolumn{1}{c|}{\cellcolor[HTML]{E6E6E6}0.980} & \cellcolor[HTML]{CCCCCC}0.990 \\ \hline
\multicolumn{1}{l}{} & \multicolumn{6}{c|}{\textbf{ArXiv}} & \multicolumn{2}{c}{\textbf{ArXiv-CS}} \\ \hline
70m & \cellcolor[HTML]{FFFFFF}0.533 & \cellcolor[HTML]{FAFAFA}0.562 & \multicolumn{1}{c|}{\cellcolor[HTML]{CECECE}0.823} & \cellcolor[HTML]{FFFFFF}0.529 & \cellcolor[HTML]{F4F4F4}0.598 & \multicolumn{1}{c|}{\cellcolor[HTML]{CFCFCF}0.820} & \multicolumn{1}{c|}{\cellcolor[HTML]{FFFFFF}0.743} & \cellcolor[HTML]{FEFEFE}0.745 \\
160M & \cellcolor[HTML]{F9F9F9}0.569 & \cellcolor[HTML]{F6F6F6}0.586 & \multicolumn{1}{c|}{\cellcolor[HTML]{D2D2D2}0.802} & \cellcolor[HTML]{FEFEFE}0.536 & \cellcolor[HTML]{F8F8F8}0.571 & \cellcolor[HTML]{D2D2D2}0.797 & \multicolumn{1}{c|}{\cellcolor[HTML]{EDEDED}0.776} & \cellcolor[HTML]{FDFDFD}0.747 \\
1.4B & \cellcolor[HTML]{FAFAFA}0.563 & \cellcolor[HTML]{F7F7F7}0.581 & \multicolumn{1}{c|}{\cellcolor[HTML]{D0D0D0}0.814} & \cellcolor[HTML]{FDFDFD}0.543 & \cellcolor[HTML]{F7F7F7}0.577 & \cellcolor[HTML]{CCCCCC}0.832 & \multicolumn{1}{c|}{\cellcolor[HTML]{E5E5E5}0.791} & \cellcolor[HTML]{DCDCDC}0.807 \\
2.8B & \cellcolor[HTML]{FFFFFF}0.532 & \cellcolor[HTML]{F8F8F8}0.570 & \multicolumn{1}{c|}{\cellcolor[HTML]{D1D1D1}0.808} & \cellcolor[HTML]{F8F8F8}0.571 & \cellcolor[HTML]{F9F9F9}0.568 & \cellcolor[HTML]{CFCFCF}0.815 & \multicolumn{1}{c|}{\cellcolor[HTML]{D3D3D3}0.824} & \cellcolor[HTML]{DCDCDC}0.807 \\
12B & \cellcolor[HTML]{F7F7F7}0.577 & \cellcolor[HTML]{F3F3F3}0.606 & \multicolumn{1}{c|}{\cellcolor[HTML]{D3D3D3}0.795} & \cellcolor[HTML]{F4F4F4}0.600 & \cellcolor[HTML]{F5F5F5}0.590 & \cellcolor[HTML]{D7D7D7}0.770 & \multicolumn{1}{c|}{\cellcolor[HTML]{CFCFCF}0.831} & \cellcolor[HTML]{CCCCCC}0.835 \\ \hline
\multicolumn{1}{l}{} & \multicolumn{6}{c|}{\textbf{Pubmed}} & \multicolumn{2}{c}{\textbf{ArXiv-Math}} \\ \hline
70m & \cellcolor[HTML]{FFFFFF}0.517 & \cellcolor[HTML]{F9F9F9}0.576 & \multicolumn{1}{c|}{\cellcolor[HTML]{D7D7D7}0.880} & \cellcolor[HTML]{FCFCFC}0.545 & \cellcolor[HTML]{F9F9F9}0.573 & \cellcolor[HTML]{D7D7D7}0.882 & \multicolumn{1}{c|}{\cellcolor[HTML]{FEFEFE}0.603} & \cellcolor[HTML]{FFFFFF}{\color[HTML]{222222} 0.601} \\
160M & \cellcolor[HTML]{FDFDFD}0.540 & \cellcolor[HTML]{F6F6F6}0.600 & \multicolumn{1}{c|}{\cellcolor[HTML]{D9D9D9}0.870} & \cellcolor[HTML]{FBFBFB}0.558 & \cellcolor[HTML]{F9F9F9}0.575 & \cellcolor[HTML]{D6D6D6}0.894 & \multicolumn{1}{c|}{\cellcolor[HTML]{F2F2F2}0.615} & \cellcolor[HTML]{FDFDFD}0.604 \\
1.4B & \cellcolor[HTML]{F8F8F8}0.583 & \cellcolor[HTML]{FBFBFB}0.558 & \multicolumn{1}{c|}{\cellcolor[HTML]{D7D7D7}0.883} & \cellcolor[HTML]{FCFCFC}0.551 & \cellcolor[HTML]{F8F8F8}0.582 & \cellcolor[HTML]{D8D8D8}0.875 & \multicolumn{1}{c|}{\cellcolor[HTML]{DDDDDD}0.637} & \cellcolor[HTML]{E1E1E1}0.633 \\
2.8B & \cellcolor[HTML]{FBFBFB}0.553 & \cellcolor[HTML]{F9F9F9}0.573 & \multicolumn{1}{c|}{\cellcolor[HTML]{DADADA}0.860} & \cellcolor[HTML]{FAFAFA}0.570 & \cellcolor[HTML]{FAFAFA}0.570 & \cellcolor[HTML]{D6D6D6}0.894 & \multicolumn{1}{c|}{\cellcolor[HTML]{E8E8E8}0.626} & \cellcolor[HTML]{D5D5D5}0.646 \\
12B & \cellcolor[HTML]{F9F9F9}0.577 & \cellcolor[HTML]{F7F7F7}0.590 & \multicolumn{1}{c|}{\cellcolor[HTML]{D5D5D5}0.900} & \cellcolor[HTML]{FAFAFA}0.570 & \cellcolor[HTML]{F9F9F9}0.580 & \cellcolor[HTML]{CCCCCC}0.980 & \multicolumn{1}{c|}{\cellcolor[HTML]{DFDFDF}0.635} & \cellcolor[HTML]{CCCCCC}0.655 \\ \hline
\end{tabular}
}
\label{tab:pythia-by-size}
\end{table}

\vspace{-.45cm}

\noindent \textbf{Unimodal.} Table \ref{tab:pythia-by-size}  shows a clear trend on the AUC as the model size grows for Pythia family. Results are similar in GPT-Neo, thus placed in Appendix B. All datasets show better results in all configurations on the 12B version, excluding ArXiv-1 month. {\color{black}
 For this dataset, both deduplicated and non-deduplicated models show improved AUC scores when scaling from 70M to 2.8B parameters. Yet, a significant decline is observed in the 12B version of the model on this dataset, with AUC values dropping from 0.92 to 0.84 (deduped) and 0.86 (non-dedup). Despite this unexpected decrease, AUC values are still very high.}

By analyzing the trends on the percentage of change of AUC of non-LP-based proposals and ours, while LPs shows an incremental trend, differences with other approaches are not significant. 


\noindent \textbf{Multimodal.} From an architectural perspective, while there are no differences in the sizes of the visual encoders, having a larger LLM on the textual+visual part affects the results.  In general, excluding again the Magpie dataset {\color{black}(since it only contains texts)}, the 7B model seems to reveal more information in both parts of the models, the visual only encoder and the textual+visual LLM, denoting higher memorization of the data than the 0.5B version.

\begin{Summary}{}{secondsummary}
As model grows, we have better AUC in the 85.9\% of the cases for both uni and multimodal LLMs. The pace is the same as non-LP-based approaches. 
\end{Summary}

\subsection{Impact of bias}
\label{sec:impact_bias}

In this case, we concentrate on unimodal LLMs since in multimodality, our results show that there are no significant differences between the average values of the member and non-member samples in the datasets based on HV or SSIM (see Appendix~A for further details). Thus, high or low AUC results seem to be influenced by task complexity or dataset nature rather than by the actual content differences. 

\medskip

\noindent \textbf{Unimodal.} LUMIA significantly outperforms \cite{liu2024probing}, which also uses LPs. Specifically, we achieve a 25\% improvement in the CS subset. These findings align with their observation that the Math subset is more challenging to predict. Nevertheless, LUMIA still achieves an AUC above 0.60. even in these more difficult subsets. Additionally, results for WikiMIA show a particularly high improvement of 41\%. On average, for TB datasets, LUMIA returns an 18\% improvement over all the previous efforts.

When studying NGB, Table \ref{tab:general} shows that LUMIA outperforms all reported configurations and baselines (except for Wikipedia Ref) across all models. They reported that no configuration reached an AUC greater than 0.60 for the {\color{black} $\mathcal{N}=13$  overlap of \ $\mathcal{P}=0.8$ on the Pythia dedup model. Contrarily, in specific cases, such as Pile-CC, DM\_math, we can reach this threshold. Additionally, for the 12B model, we achieve an AUC of 0.58 on PubMed and 0.584 on Hackernews, both approaching the 0.60 threshold more closely than previous approaches.}
All in all, results for $\mathcal{N}=13$ with  $\mathcal{P}=0.8$ lead to an overall improvement of 13.10\%. 

{\color{black}For $\mathcal{N}=7$ with $\mathcal{P}=0.2$ } on the Pythia 12B family in Table \ref{tab:general}, our approach consistently outperforms the state of the art, with improvements ranging from a minimum of 2\% on Wikipedia to a maximum of 64\% on PubMed and an average of 18.74\%. {\color{black} Notably, on the DM\_math dataset the ref method performs poorly under overlap configurations of $\mathcal{P}=0.2$ with $\mathcal{N}=13$ and $\mathcal{N}=7$, achieving AUC scores of 0.44 and 0.41, respectively. Consequently, our approach surpasses the ref method  by 68\% and 129\%.
}

For GPT-Neo 2.7B, in line with the previous model, all configurations overtake results from Duan \textit{et al.} \cite{duan2024llms} with an overall improvement of  14.43\%. Nonetheless, in this case of {\color{black} $\mathcal{N}=13$  overlap with $\mathcal{P}=0.8$ }, none of our configurations, excluding Github, overtakes the 0.6 AUC. 

Although the distinction between members and non members are based on different techniques, it is generally observed that TB datasets are easier to detect than NGB ones, even in cases of high overlap. For instance, the Wikipedia dataset in the NGB dataset achieves a maximum AUC of 0.685, while the Temporal-Wiki dataset in the TB datasets reaches up to 0.95 AUC.

\begin{Summary}{}{thirdsummary}
\textit{Unimodal:} {\color{black} In line with non-LPs approaches,} as overlap gets reduced, we have better results in all cases. TB datasets present better results than NGB and all results are above 0.6.

\textit{Multimodal:} HV and SSIM differences show no correlation with AUC, suggesting that the results are more influenced by task or dataset complexity rather than any bias between member and non-member samples.

\end{Summary}

\subsection{Impact of dataset nature}
 \label{sec:dataNature}

\noindent \textbf{Unimodal.} Table \ref{tab:compProbes} shows consistent conclusions with those of Liu \textit{et al.} \cite{liu2024probing} on TB datasets, who argue that the difficulty of the content impacts results. For example, our approach overtakes their results on a 25\% and 12\% on the arXiv-CS and arXiv-Math datasets respectively, but in line with their hypothesis, our LPs also perform worse on the arXiv-Math dataset, {\color{black} where the nature of the text content makes detection more challenging as mathematical texts are more complex and harder to memorize by the LLMs.}

In the case of NGB datasets, as shown in Table \ref{tab:general}, similar patterns are observed. For instance, on Github,  which was the easiest to predict on the $\mathcal{N}=13$  with $\mathcal{P}=0.8$ overlap according to Duan \textit{et al.} \cite{duan2024llms}, LUMIA also offers the best results. Code-related samples may contain HTML tags and unique variable names, which could make the members and non-members more identifiable. Furthermore, other datasets such as Wikipedia or Hackernews, which contain a wider range of topics and variety of texts, make harder the identification of differences between members and non-members.

\medskip

\noindent \textbf{Multimodal.} In multimodal datasets, the type of information  impacts the results. Table \ref{tab:compmulti} shows that, except for ChartQA and IconQA, the model appears to add more information through the visual encoders, particularly with images. For example, in the case of Magpie, which is a text-only dataset, the visual encoder returns an almost random AUC of 0.52, but it adds more information when dealing with both textual and visual inputs, reaching a 0.572 AUC.

{
\color{black}
For Textcaps, the prompt remains the same for both members and non-members, while the images exhibit greater variability. This setup results in a slight drop in accuracy when using the combined textual and visual parts of the model, with AUC decreasing from 0.617 for visual-only LPs to 0.601 for visual+text LPs. The consistent prompt across members and non-members likely introduces noise, diminishing the model's ability to differentiate between them.}

{
\color{black} Finally, datasets that follow a consistent template across the prompts of both members and non-members, such as MathV360k, demonstrate a reduced ability for classifiers to distinguish between classes compared to datasets with more varied images and texts. For example, datasets like ScienceQA and IconQA, which lack a uniform template across samples, achieve AUC values around 0.8 for both Textual+Visual and Visual-only configurations. 

}

\begin{Summary}{}{fourthdsummary}
\textit{Unimodal:} Code or datasets containing mathematical formulas are easier to identify than general-purpose texts. LPs are specially good on detecting these modalities.

\textit{Multimodal:} Repetitive prompts strengthen resistance to MIA, while a greater variety of images makes models more vulnerable.

\end{Summary}

\subsection{Impact of deduplication}
\label{sec:dataDedup}

Since data in Llava and OneVision models is already deduplicated, and no open-source models without this data processing exist, only unimodality is considered.

\medskip

\noindent \textbf{Unimodal.} Results focus on Pythia, Table \ref{tab:pythia-by-size}, since it is the only model which provides a clear distinction of deduplicated data. For the TB datasets, MIAs tend to be more effective on non-deduplicated models. This is likely because deduplication reduces the repetition of data in the training set, thereby limiting the model’s ability to memorize and overfit to specific patterns. 

In contrast, for the NGB datasets, no significant differences are observed between the deduplicated and non-deduplicated versions of the models.

\subsection{Analysis per model and layer}
\label{sec:layerDep}
\label{sec:permodelayerAUC}

\begin{figure*}[!hbt]
 \centering
 \subfigure[]{
    \includegraphics[width=0.41\textwidth]{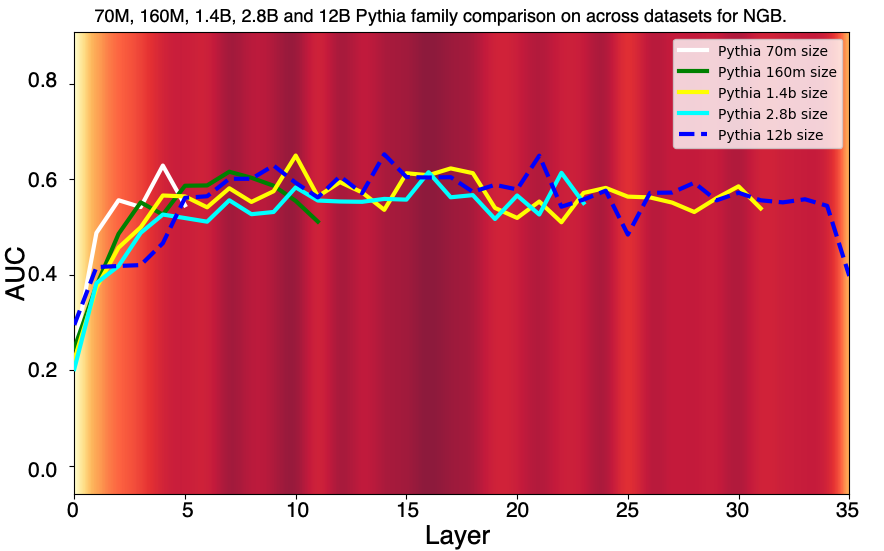}
 } 
 \subfigure[]{
    \includegraphics[width=0.41\textwidth]{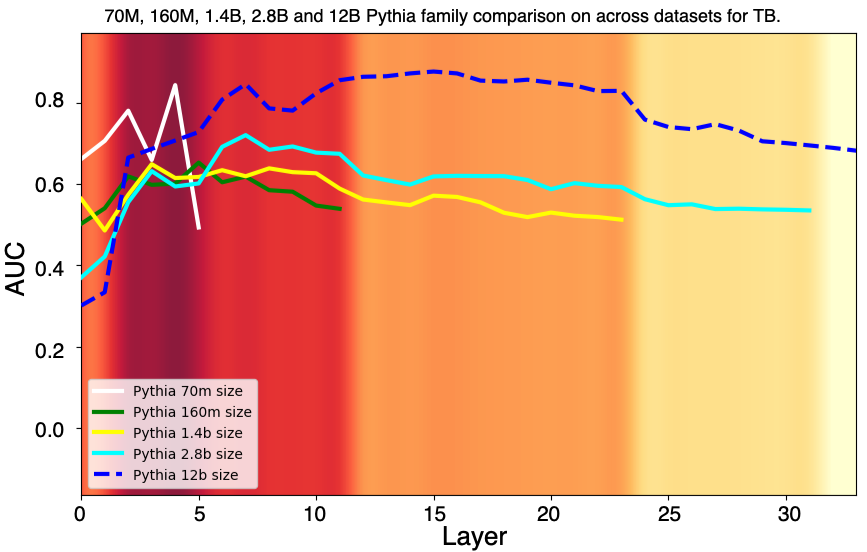}
 }
 \caption{Pythia family. AUC by layer. (a) NGB datasets; (b) TB datasets}
\label{fig:comparisonlayers}
\end{figure*}

\begin{figure*}[!hbt]
  \centering
  \subfigure[]{
    \includegraphics[width=0.41\textwidth]{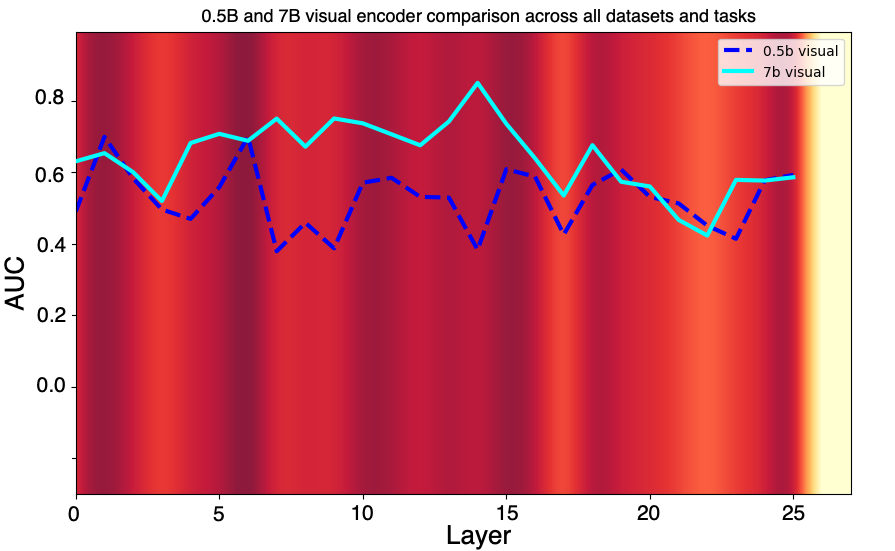}
  } 
  \subfigure[]{
    \includegraphics[width=0.41\textwidth]{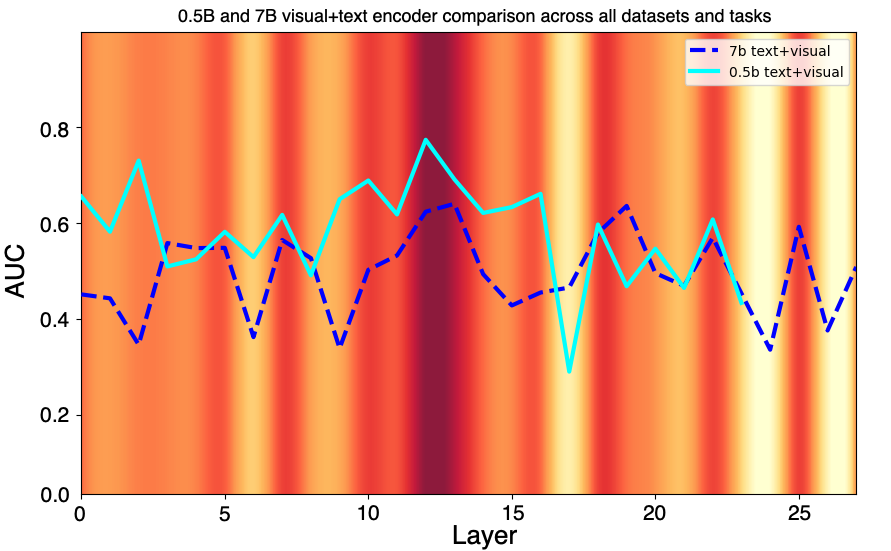}
  }
  \caption{LLava-OneVision. AUC by layer. (a) Visual encoder only; (b) Visual+text encoder}
 
\label{fig:comparisonlayersMulti}
\end{figure*}

Figures \ref{fig:comparisonlayers} and \ref{fig:comparisonlayersMulti} present results from Pythia and the multimodal models respectively. They include the normalized average values of the AUC across all datasets. Gradient colors represent the average AUC for each layer, calculated across all models and datasets. Results for GPT-Neo are omitted for brevity, as they are always more effective on deeper layers (see Appendix B for details). 

\vspace{-1.5cm}

\noindent \textbf{Unimodal.} Figure \ref{fig:comparisonlayers} illustrates the AUC across layers for the Pythia model family, covering results from all datasets and model types (both deduplicated and non-deduplicated). In Figure \ref{fig:comparisonlayers}(a), model performance on all NGB datasets from Mimir are reported. Notably, there are certain layers where model performance peaks: particularly around layer 10 and again between layers 15 and 18, where larger models achieve higher AUC values. This suggests that specific depths of the model add more information useful for LPs to detect MIAs.

In Figure \ref{fig:comparisonlayers}(b), the average normalized values for the same models on the TB datasets are reported. Peak performance begins around layers 3-5, showing that earlier layers on the model are enough to get good results. However for larger models, between layers 5 and 12, there is also a portion of the model where important information for the membership inference is revealed.

\noindent \textbf{Multimodal.} Figure \ref{fig:comparisonlayersMulti} shows results for the multimodal model. Figure \ref{fig:comparisonlayersMulti}(a) presents the AUC for the visual encoder, which, despite differences in scale, reveals common areas around the layer 15, which may denote that deeper layers add more information.

In Figure \ref{fig:comparisonlayersMulti}(b), similarly, the AUC by layer is shown for the visual+text encoder. It is particularly noteworthy that around layers 12-13 both modalities, that is visual and visual+text, exhibit a spike in information useful for MIA.

\begin{Summary}{}{sixthsummary}

\textit{Unimodal:}  LPs on Pythia family with TB datasets are more effective on earlier layers (4-5). For NGB datasets,  deeper layers (15-16) are preferred. LPs in GPT-Neo are always more effective on deeper layers (25)

\textit{Multimodal:}  Middle layers reveal more information in visual (8-9 or 15) as well as in textual+visual part (13-14).

\end{Summary}

\section{Related work}
\label{RW}

MIA attacks have been largely studied. 
From a grey-box perspective, Shi \textit{et al.} \cite{shi2023detecting} analyze output logits of LLMs based on the assumption that unseen samples tend to contain outlier words with very low probabilities.

Meeus \textit{et al.} \cite{meeus2024did} adopt a binary classifier to distinguish members from non-members using document-level features and a normalization algorithm in a black-box approach.
 Although promising, Das \textit{et al.} \cite{das2024blind} achieve superior results than other works. They hypothesize that good results can be achieved by leveraging heuristics based on the features and statistics of public MIA datasets.

Duan \textit{et al.} \cite{duan2024membership} introduce some benchmark datasets, \textit{Mimir}, designed to address potential biases and assess state-of-the-art MIA methods. 
They also show that cutoff dates are crucial, as the overlap of N-grams may fluctuate over time.
 A similar work is proposed by 
Kim \textit{et al.} \cite{kim2024detecting}. They introduce a new maximization expectation algorithm. Nevertheless they highlight their results are close to random guessing when the distribution of member and non-members are close. 

A comparable white-box approach is presented by Liu \textit{et al.} \cite{liu2024probing}, who use simple linear classifiers on the activations of the LLMs. They fine-tune a pretrained model 
using a prompt to ensure that members and non-members are represented in a standardized format. 
Moreover, they consider only layer \textit{l}, thus neglecting a per layer analysis of the results.

In terms of multimodality in LLMs, Li \textit{\textit{et al.}} \cite{li2024membership} proposed a grey-box MIA approach.
This approach introduces a novel metric that relies on the confidence level of the model's output. As it relies on generating long sequences of text as output, this approach may have limited generalizability. Moreover, results are worse than those of LUMIA.

\begin{table}
  \centering
  \caption{Related work analysis}
    \scalebox{0.62}{
\begin{tabular}{|p{2cm}|p{4.5cm}|p{2.3cm}|p{2.3cm}|p{1.2cm}|p{1.2cm}|p{1.2cm}|p{1.7cm}|p{1.5cm}|}
\hline
\textbf{Ref.} & \textbf{Dataset} & \textbf{Model} & \textbf{Input features} & \textbf{Multi-modal-ity} & \textbf{Per layer analysis} & \textbf{Dedup/ Non- dedup} & \textbf{Temporal Bias/ ngram bias Analysis} & \textbf{Whitebox (W) / Greybox (G) / Blackbox (B)} \\ \hline
Duan \textit{\textit{et al.}} \cite{duan2024membership} & Mimir, Temporal wiki, Temporal ArXiv & Pythia, Pythia-dedup, GPT-Neo & Loss from models' logits & $\times$ & $\times$ & $\checkmark$ & $\checkmark$ & G \\ \hline
Shi \textit{et al.}\cite{shi2023detecting} & WikiMIA, BookMia & Pythia, GPT-Neo, Llama, OPT &  & $\times$ & $\times$ & $\times$ & $\times$ & G \\ \hline
Das \textit{et al.} \cite{das2024blind} & WikiMIA, BookMia, Temporal-wiki, temporal-ArXiv, ArXiv-1 month, Gutenberg & - & Features from the texts & $\times$ & $\times$ & $\times$ & $\checkmark$ & B \\ \hline
Meeus \textit{et al.} \cite{meeus2024did} & Gutenberg, ArXiv papers & Open-Llama & Features from the texts & $\times$ & $\times$ & $\times$ & $\times$ & B \\ \hline
Kim \textit{et al.} \cite{kim2024detecting} & WikiMIA, OLMoMIA & Mamba, Pythia, Llama, OPT, GPT-Neo & Texts and membership scores & $\times$ & $\times$ & $\times$ & $\times$ & B \\ \hline
Li \textit{et al.} \cite{li2024membership} & VL-MIA & LLaVA 1.5, MiniGPT-4,LLaMA\_adapter v2 & Instruction based on the image, prompt and the output of the model of previous prompt & $\checkmark$ & $\times$ & $\times$ & $\times$ & G \\ \hline
Liu \textit{et al.} \cite{liu2024probing} & ArXivMIA, WikiMIA & Pythia, OPT, Tiny-Llama, Open-Llama & Activations & $\times$ & $\times$ & $\times$ & $\times$ & W \\ \hline
\textbf{LUMIA} & \textbf{WikiMIA, ArXiv-MIA, Temporal ArXiv/wik, ArXiv-1-month, Gutenberg, Mimir (The Pile), Textcaps, MathV360, AOK,  ChartQA, ScienceQA, IconQA, Magpie} & \textbf{Pythia, Pythia-dedup, GPT-Neo, LLava-OneVision} & \textbf{Activations} & \textbf{$\checkmark$} & \textbf{$\checkmark$} & \textbf{$\checkmark$} & \textbf{$\checkmark$} & \textbf{W} \\ \hline
\end{tabular}}
\end{table}

All in all, Table \ref{sec:conclusion} presents an overview of related works together with LUMIA. Our method is assessed on a broader range of datasets, covering both TB and NGB datasets, as well as deduplicated and non-deduplicated models—a distinction only addressed by Duan \textit{\textit{et al.}} \cite{duan2024llms}. Furthermore, LUMIA is the only study to conduct a layer-by-layer analysis, which provides valuable insights into how unimodal and multimodal LLMs process information. Lastly, to the best of authors' knowledge, our study is the only one that examines the impact of data type on MIAs within a multimodal context.

\section{Conclusion}
\label{sec:conclusion}
In this paper, an approach (dubbed LUMIA) has been proposed to tackle Membership Inference Attacks (MIAs). LUMIA helps on determining whether a sample was used during the pre-training of a target model. Remarkably, LUMIA leverages Linear Probes, thus adopting a white-box approach. LUMIA has been tested on a wide range of datasets and different LLMs, both for uni- and multimodal cases. Our results show that it overtakes the state of the art, maintaining consistency across datasets, regardless of the presence or absence of bias.  

As future work, LUMIA could be extended to other modalities, such as video or audio, along with exploring its applicability in detecting copyright violations. Additionally, two key future directions are devised, namely, leveraging insights about specific layers to introduce noise into those revealing the most information, enhancing the model's resilience to such attacks, and conducting a deeper analysis of these layers to optimize the results.

\appendix

\section*{Appendices}

\subsection*{A. Bias analysis on multimodal models.}
\label{ref:Bias_analysis}

Table~\ref{tab:statsmulti} summarizes the statistics for all datasets containing images in the multimodal case, where the right column shows the absolute difference of the values between members and non-members. In further detail, content-related differences are analyzed from a statistical perspective, with both HV and SSIM computed. HV provides insights into the variation in image content, while SSIM quantifies structural similarity. Together, these measures help to understand content-driven distinctions between members and non-members.

HV shows no clear correlation with AUC values, as seen in Table~\ref{tab:compmulti}. For instance, the ScienceQA dataset achieves a high AUC of 0.99 despite a large HV difference of 0.80, while IconQA, with a much smaller HV difference of 0.10, still achieves a strong AUC of 0.869. A similar pattern emerges with SSIM. For example, ScienceQA, with an SSIM difference of 0.08, achieves an AUC of 0.99, whereas Mathv360k, with the same SSIM difference, only reaches 0.66 AUC. Thus, high or low AUC results seem to be influenced by task complexity or dataset nature rather than by the actual content differences among member and non-member samples.

\begin{table}[!ht]
\caption{Multimodal datasets. Bias analysis using Hash variation (HV) and Structural similarity Index (SSIM) in \%.}
\centering
\begin{tabular}{c|cc|cc|cc}
\hline
\multicolumn{1}{l|}{} & \multicolumn{2}{c|}{\textbf{Members}} & \multicolumn{2}{c|}{\textbf{Non members}} & \multicolumn{2}{c}{\textbf{Difference (abs.)}} \\ \hline
\textbf{Dataset} & \textbf{HV} & \textbf{SSIM} & \textbf{HV} & \textbf{SSIM} & \textbf{HV} & \textbf{SSIM} \\ \hline
AOK & 30.53 & 0.69 & 30.75 & 0.72 & 0.22 & 0.03 \\
Textcaps & 3.52 & 0.67 & 3.82 & 0.68 & 0.30 & 0.01 \\
ScienceQA & 30.36 & 0.62 & 29.56 & 0.54 & 0.80 & 0.08 \\
ChartQA & 28.19 & 0.26 & 28.33 & 0.3 & 0.14 & 0.04 \\
IconQA & 30.94 & 0.53 & 31.04 & 0.56 & 0.10 & 0.03 \\
Mathv360k & 31.08 & 0.54 & 30.9 & 0.46 & 0.18 & 0.08 \\ \hline
\end{tabular}
\label{tab:statsmulti}
\end{table}

\subsection*{B. GPT-Neo. Analysis per layer and model size}
\begin{figure*}[!hbt]
  \centering
   \subfigure[]{
    \includegraphics[width=0.41\textwidth]{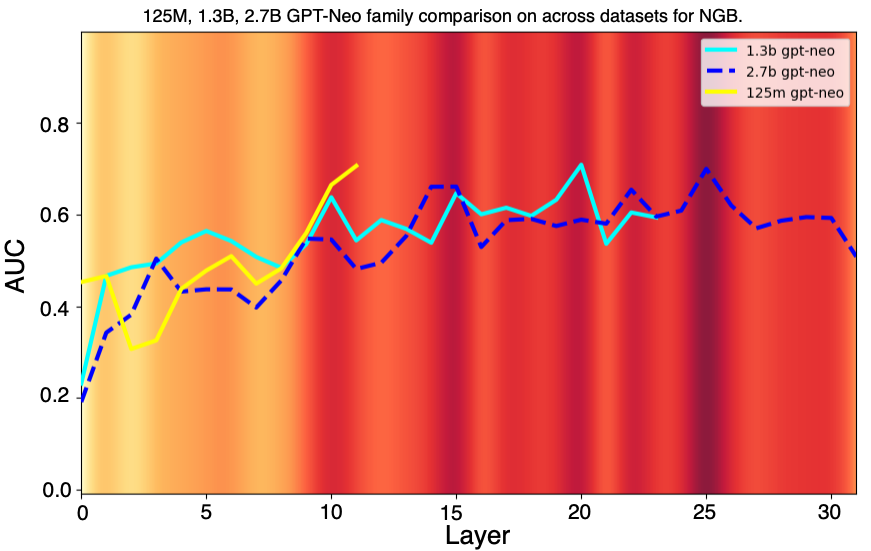}
   } 
   \subfigure[]{
    \includegraphics[width=0.41\textwidth]{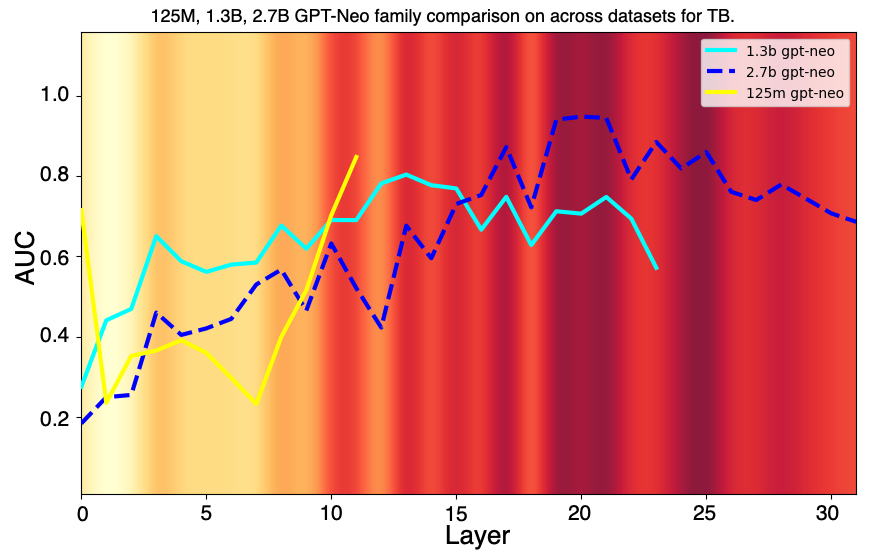}
   }
  \caption{GPT-Neo. AUC by layer. (a) NGB datasets; (b) TB datasets}
 
\label{fig:comparisonlayersgptneo}
\end{figure*}

Figure \ref{fig:comparisonlayersgptneo} presents results of the analysis of AUC per layer.  there are common areas of better performance between NGB and TB in subfigures (a) and (b), respectively. In particular, layers 10 and 25 show areas where more information useful for classifiers is added.

In what comes to the impact of model size, the same trend as in Pythia is noticed in Table \ref{tab:gpt-by-size} -- as the model grows, better AUC is achieved. The only exception is arXiv-CS, were AUC of the largest model is 0.802, while the configuration 1.3B of parameters gets 0.842.

\begin{table}[!h]
\centering
\caption{GPT-Neo. AUC per model size.}
\resizebox{0.65\columnwidth}{!}{%
\begin{tabular}{ccccc}
\hline
\multicolumn{1}{l}{} &  \multicolumn{1}{c}{\textbf{}}& \multicolumn{1}{c}{\textbf{NGB}}  &  \multicolumn{1}{c}{\textbf{}}& \multicolumn{1}{|c}{\textbf{TB}}\\ \hline
\multicolumn{1}{l}{} & \multicolumn{4}{c}{\textbf{GPT-Neo}} \\ \hline

\multicolumn{1}{l}{} & \multicolumn{3}{c|}{\textbf{Wikipedia}} &  \\ \cline{1-4}
\textbf{Params} & \textbf{$\mathcal{N}=13$  $\mathcal{P}=0.8$} & \multicolumn{1}{c}{\textbf{$\mathcal{N}=13$  $\mathcal{P}=0.2$}} & \multicolumn{1}{c|}{\textbf{$\mathcal{N}=7$  $\mathcal{P}=0.2$}} & \multirow{-2}{*}{\textbf{Gutenberg}} \\ \hline
125M & \multicolumn{1}{c}{\cellcolor[HTML]{ECECEC}0.590} & \cellcolor[HTML]{FFFFFF}0.530 & \multicolumn{1}{c|}{\cellcolor[HTML]{ECECEC}0.590} & \cellcolor[HTML]{FFFFFF}0.940 \\
1.3B & \multicolumn{1}{c}{\cellcolor[HTML]{ECECEC}0.590} & \cellcolor[HTML]{FCFCFC}0.560 & \multicolumn{1}{c|}{\cellcolor[HTML]{CCCCCC}0.650} & \cellcolor[HTML]{CCCCCC}0.970 \\
2.7B & \multicolumn{1}{c}{\cellcolor[HTML]{ECECEC}0.590} & \cellcolor[HTML]{F7F7F7}0.570 & \multicolumn{1}{c|}{\cellcolor[HTML]{CCCCCC}0.650} & \cellcolor[HTML]{CCCCCC}0.970 \\ \hline
\multicolumn{1}{l}{} & \multicolumn{3}{c|}{\textbf{Github}} & \textbf{arXiv-1 month} \\ \hline
125M & \multicolumn{1}{c}{\cellcolor[HTML]{FFFFFF}0.650} & \cellcolor[HTML]{E0E0E0}0.830 & \multicolumn{1}{c|}{\cellcolor[HTML]{D2D2D2}0.910} & \cellcolor[HTML]{FFFFFF}0.820 \\
1.3B & \multicolumn{1}{c}{\cellcolor[HTML]{E9E9E9}0.780} & \cellcolor[HTML]{E0E0E0}0.830 & \multicolumn{1}{c|}{\cellcolor[HTML]{CCCCCC}0.940} & \cellcolor[HTML]{E0E0E0}0.910 \\
2.7B & \multicolumn{1}{c}{\cellcolor[HTML]{EAEAEA}0.770} & \cellcolor[HTML]{DCDCDC}0.850 & \multicolumn{1}{c|}{\cellcolor[HTML]{CCCCCC}0.940} & \cellcolor[HTML]{D9D9D9}0.930 \\ \hline
\multicolumn{1}{l}{} & \multicolumn{3}{c|}{\textbf{Pile CC}} & \textbf{Temporal wiki} \\ \hline
125M & \multicolumn{1}{c}{\cellcolor[HTML]{FFFFFF}0.550} & \cellcolor[HTML]{F9F9F9}0.560 & \multicolumn{1}{c|}{\cellcolor[HTML]{F3F3F3}0.570} & \cellcolor[HTML]{FFFFFF}0.887 \\
1.3B & \multicolumn{1}{c}{\cellcolor[HTML]{E6E6E6}0.590} & \cellcolor[HTML]{F3F3F3}0.570 & \multicolumn{1}{c|}{\cellcolor[HTML]{E6E6E6}0.590} & \cellcolor[HTML]{DEDEDE}0.908 \\
2.7B & \multicolumn{1}{c}{\cellcolor[HTML]{F9F9F9}0.560} & \cellcolor[HTML]{E6E6E6}0.590 & \multicolumn{1}{c|}{\cellcolor[HTML]{CCCCCC}0.630} & \cellcolor[HTML]{CCCCCC}0.919 \\ \hline
\multicolumn{1}{l}{} & \multicolumn{3}{c|}{\textbf{DM Math}} & \textbf{Temporal arXiv} \\ \hline
125M & \multicolumn{1}{c}{\cellcolor[HTML]{FFFFFF}0.550} & \cellcolor[HTML]{F0F0F0}0.690 & \multicolumn{1}{c|}{\cellcolor[HTML]{CCCCCC}1.000} & \cellcolor[HTML]{FFFFFF}0.728 \\
1.3B & \multicolumn{1}{c}{\cellcolor[HTML]{FEFEFE}0.560} & \cellcolor[HTML]{EBEBEB}0.730 & \multicolumn{1}{c|}{\cellcolor[HTML]{CCCCCC}1.000} & \cellcolor[HTML]{D4D4D4}0.777 \\
2.7B & \multicolumn{1}{c}{\cellcolor[HTML]{FDFDFD}0.570} & \cellcolor[HTML]{E9E9E9}0.750 & \multicolumn{1}{c|}{\cellcolor[HTML]{CCCCCC}1.000} & \cellcolor[HTML]{CCCCCC}0.786 \\ \hline
\multicolumn{1}{l}{} & \multicolumn{3}{c|}{\textbf{Hackernews}} & \textbf{WikiMIA} \\ \hline
125M & \multicolumn{1}{c}{\cellcolor[HTML]{FFFFFF}0.550} & \cellcolor[HTML]{FFFFFF}0.550 & \multicolumn{1}{c|}{\cellcolor[HTML]{F1F1F1}0.570} & \cellcolor[HTML]{FFFFFF}0.897 \\
1.3B & \multicolumn{1}{c}{\cellcolor[HTML]{EAEAEA}0.580} & \cellcolor[HTML]{F8F8F8}0.560 & \multicolumn{1}{c|}{\cellcolor[HTML]{E2E2E2}0.590} & \cellcolor[HTML]{CECECE}0.985 \\
2.7B & \multicolumn{1}{c}{\cellcolor[HTML]{E2E2E2}0.590} & \cellcolor[HTML]{DBDBDB}0.600 & \multicolumn{1}{c|}{\cellcolor[HTML]{CCCCCC}0.620} & \cellcolor[HTML]{CCCCCC}0.987 \\ \hline
\multicolumn{1}{l}{} & \multicolumn{3}{c|}{\textbf{Arxiv}} & \textbf{arXiv-CS} \\ \hline
125M & \cellcolor[HTML]{FFFFFF}0.550 & \cellcolor[HTML]{FEFEFE}0.560 & \multicolumn{1}{c|}{\cellcolor[HTML]{D2D2D2}0.820} & \cellcolor[HTML]{FFFFFF}0.681 \\
1.3B & \cellcolor[HTML]{FCFCFC}0.570 & \cellcolor[HTML]{FCFCFC}0.570 & \multicolumn{1}{c|}{\cellcolor[HTML]{CCCCCC}0.850} & \cellcolor[HTML]{CCCCCC}0.842 \\
2.7B & \cellcolor[HTML]{FAFAFA}0.580 & \cellcolor[HTML]{F9F9F9}0.590 & \multicolumn{1}{c|}{\cellcolor[HTML]{CCCCCC}0.850} & \cellcolor[HTML]{D9D9D9}0.802 \\ \hline
\multicolumn{1}{l}{} & \multicolumn{3}{c|}{\textbf{Pubmed}} & \textbf{arXiv-Math} \\ \hline
125M & \cellcolor[HTML]{FFFFFF}0.540 & \cellcolor[HTML]{FFFFFF}0.540 & \multicolumn{1}{c|}{\cellcolor[HTML]{D9D9D9}0.830} & \cellcolor[HTML]{FFFFFF}0.604 \\
1.3B & \cellcolor[HTML]{FDFDFD}0.560 & \cellcolor[HTML]{FFFFFF}0.540 & \multicolumn{1}{c|}{\cellcolor[HTML]{CCCCCC}0.920} & \cellcolor[HTML]{DCDCDC}0.633 \\
2.7B & \cellcolor[HTML]{FAFAFA}0.580 & \cellcolor[HTML]{FBFBFB}0.570 & \multicolumn{1}{c|}{\cellcolor[HTML]{CCCCCC}0.920} & \cellcolor[HTML]{CCCCCC}0.646 \\ \hline
\end{tabular}
\label{tab:gpt-by-size}
}
\end{table}

\subsection*{C. Acknowledgements}
Nicolas Anciaux was supported by the French grant \href{https://www.pepr-cybersecurite.fr\\/projet/ipop/}{iPoP} PEPR (ANR-22-PECY-0002). Luis Ibanez-Lissen was supported by the Spanish National Cybersecurity Institute (INCIBE) grant APAMciber within the framework of the Recovery, Transformation and Resilience Plan funds, financed by the European Union (Next Generation). Jose Maria de Fuentes and also Lorena Gonzalez partially supported by grant PID2023-150310OB-I00 of the Spanish AEI.

\bibliographystyle{splncs04}
\bibliography{sample-base}
\end{document}